# Harmonised benchmarking of foundation models for single-cell and spatial transcriptomics reveals context-dependent generalisation

Sally Chen[1,2], Roxana Zahedi[1], Lucy Chhuo[1], Ricky Nguyen[1,2], Marjan BaghGolshani[3], Amin Beheshti[3], Mark Grosser[4], Min Yang[5], Nona Farbehi[6], Nigel Lovell[6], Ahmadreza Argha[6], Thantrira Porntaveetusm[7], Fatemeh Vafaee[2,8], Youqiong Ye[*9], Hamid Alinejad-Rokny[*1,10]

[1] BioMedical Machine Learning Laboratory, School of Biomedical Engineering, UNSW Sydney, Randwick, 2032, Australia

[2] School of Biotechnology and Biomolecular Sciences, UNSW Sydney, Randwick, 2032, Australia

[3] School of Computing, Macquarie University, Sydney, NSW 2109, Australia

[4] 23Strands, Pyrmont, Australia

[5] Shenzhen Key Laboratory for High Performance Data Mining, Shenzhen Institute of Advanced Technology, Shenzhen, China

[6] School of Biomedical Engineering, UNSW Sydney, Randwick, 2032, Australia

[7] Center of Excellence in Precision Medicine and Digital Health, Faculty of Dentistry, Chulalongkorn University, Bangkok, Thailand

[8] UNSW AI Institute, University of New South Wales, Sydney, NSW 2032, Australia

[9] Shanghai Institute of Immunology, Shanghai Jiao Tong University School of Medicine, Shanghai 200025, China

[10] Visiting Scholar (Collaborative Projects), Center of Excellence in Precision Medicine and Digital Health, Chulalongkorn University, Bangkok, Thailand

* Correspondence should be addressed to H.A.R. (h.alinejad@unsw.edu.au) and Y.Y. (youqiong.ye@shsmu.edu.cn)

**Abstract**

Single-cell and spatial foundation models promise transferable biological representations, yet their generality remains largely untested across modalities, biological domains and analytical tasks. We benchmarked six representative models, Nicheformer, CellPLM, scGPT-spatial, GenePT, scELMo and Novae, using a harmonised framework spanning scRNA-seq, spatial transcriptomics and Perturb-seq. We evaluated zero-shot and continually pretrained clustering, supervised annotation, marker-gene concordance and perturbation prediction. Model performance was strongly conditional: expression-trained cell-level transformers best resolved many cell-identity tasks, spatial and graph-aware models better preserved tissue architecture, and language-derived gene embeddings were competitive for selected perturbation-response metrics. No model dominated across tasks, and rankings shifted with modality, preprocessing, tokenisation, biological prior, domain shift and metric choice. This benchmark provides practical guidance for model selection and argues that future models should be judged by biological generalisation, interpretability and perturbation-grounded validity, not by scale or leaderboard performance alone.

## Introduction

Single-cell RNA sequencing (scRNA-seq) has transformed the study of cellular identity, developmental trajectories and disease-associated cell states[1,2]. Spatial transcriptomics (SRT) extends this view by measuring gene expression in situ, preserving tissue architecture and spatially organised cell–cell interactions[3-6]. Together, these technologies provide complementary views of cellular systems, from dissociated cell states to spatially structured tissue niches. However, their analysis remains challenging because datasets differ in platform, resolution, species, tissue context and technical noise. As a result, common tasks such as clustering, cell type annotation, batch correction, data integration, spatial domain discovery and perturbation prediction often remain sensitive to preprocessing choices and dataset-specific assumptions.

Foundation models (FMs) have emerged as a promising strategy for learning reusable representations across single-cell and spatial datasets[7,8]. Models such as scGPT, GenePT and Geneformer use gene-level, text-derived or rank-based tokenisation strategies to encode transcriptional states and support transfer across downstream tasks[9-11]. Other models, including CellPLM and Nicheformer, extend this paradigm by learning cell-level or microenvironment-aware representations from large-scale single-cell and spatial omics corpora[12,13]. These architectural choices are biologically consequential: they determine whether a model primarily captures gene-level semantics, cell-state continua, spatial neighbourhoods, tissue domains or perturbational responses.

Despite rapid progress, it remains unclear when foundation models improve biological inference in practice. Existing evaluations are often restricted to individual models, single modalities or narrow tasks such as perturbation prediction, label transfer or cell type annotation[14-16]. Recent zero-shot studies further show that widely used single-cell foundation models can underperform simpler baselines and degrade under dataset or platform shift, challenging assumptions of broad out-of-distribution generalisation[17]. Current benchmarks therefore provide limited guidance on which models generalise across both scRNA-seq and SRT, which design choices are most robust, and whether learned embeddings preserve biologically meaningful structure rather than dataset-specific correlations.

Here, we present a harmonised benchmark of six foundation models for single-cell and spatial omics: Nicheformer, CellPLM, scGPT-spatial, GenePT, scELMo and Novae. These models span gene-level, cell-level, language-derived, graph-based and spatially informed architectures. We evaluate them across diverse scRNA-seq and SRT datasets using harmonised evaluation framework with model-compatible preprocessing, shared metrics and multiple downstream tasks, including zero-shot and continually pretrained clustering, supervised cell type annotation, differential-expression-based biological validation and perturbation prediction. Our results show that preprocessing choices, tokenisation strategies and biological priors substantially influence model rankings, and that current foundation models remain vulnerable to domain shift and limited interpretability. This benchmark provides a reproducible framework for model selection and establishes practical design principles for future foundation models in single-cell and spatial omics. **Figure 1** outlines the conceptual framework of our benchmark, highlighting how model architecture, pretraining strategy, modality and downstream task were systematically compared across single-cell and spatial omics**.**

## Results

### An overview of the foundation models used in this study

We benchmarked six foundation models selected to represent major design strategies in single-cell and spatial omics rather than variants of a single modelling family (**Table 1**). GenePT[11] and scELMo[18] test whether biological knowledge encoded in natural-language descriptions of genes and cell states can produce useful cellular representations with limited or no expression-scale pretraining. CellPLM[12] and scGPT-spatial, which builds on the scGPT framework[9,19] represent expression-trained transformer models that learn cellular context from large transcriptomic corpora; scGPT-spatial further evaluates whether continued exposure to spatial transcriptomics improves tissue-aware representations. Nicheformer[13] and Novae[20] extend the comparison toward spatially resolved modelling: Nicheformer leverages joint single-cell and spatial pretraining to encode cells in microenvironmental context, whereas Novae uses graph-based self-supervision to learn tissue domains from local neighbourhood structure.

Models were selected according to the scope of the benchmark rather than by attempting to include all single-cell analysis tools. Inclusion required that a method provide reusable gene-, cell- or tissue-level representations, be applicable to scRNA-seq and/or spatial transcriptomics, and represent a distinct foundation-model design principle. We therefore excluded methods primarily developed for synthetic data generation, statistical simulation, regulatory-network inference or perturbation-only prediction from the primary model panel (**Supplementary Table S1**). This distinction allowed us to compare foundation-model representations under shared downstream tasks, rather than conflating general-purpose representation models with specialised simulators, mechanistic tools or task-specific predictors.

This model panel was designed to test a central question for the field: whether different foundation model assumptions lead to different forms of biological generalisation. Gene-centric and language-derived models may better preserve functional gene relationships, expression-trained transformers may better resolve transcriptional cell states, and spatial or graph-based models may be better suited to tissue domains and niches. Evaluating these models under a shared preprocessing and evaluation framework therefore allowed us to separate effects that are often conflated in model-specific studies, including the source and scale of pretraining data, the biological unit used for tokenisation, and the inductive bias used to represent spatial context. These contrasts motivate the downstream benchmark across zero-shot clustering, continual pretraining, supervised annotation, marker-gene concordance and perturbation prediction. **Table 1** shows an overview of the models included in the benchmark, detailing the tokenisation strategy, pretraining datasets, modalities, architecture, spatial awareness, reported tasks, pretraining objective and publication.

### Zero-shot embeddings recover biological structure but generalise unevenly across modalities

Zero-shot clustering provides a stringent test of whether pretrained foundation model embeddings encode biologically meaningful structure before exposure to dataset-specific labels or task-specific training. We therefore extracted embeddings from each applicable pretrained model and applied a matched Leiden clustering workflow, using a fixed random seed and tuning the resolution so that the number of clusters matched the number of annotated classes or domains. Performance was evaluated using ARI, NMI and silhouette score (SIL) across three scRNA-seq datasets and three spatial transcriptomics datasets (**Figure 2a, 2b**; **Table 2**). This design allowed us to test two related but distinct forms of generalisation: recovery of transcriptional cell identities in dissociated single-cell data, and recovery of anatomical or pathological tissue domains in spatial data.

In the scRNA-seq datasets, zero-shot performance depended strongly on the biological granularity of the labels. In ATAA, where smooth muscle cells, fibroblasts, endothelial cells and immune populations are separated by robust transcriptional programmes, expression-trained transformer models showed the clearest organisation. scGPT-spatial achieved the highest concordance with expert annotations (ARI = 0.87, NMI = 0.85, SIL = 0.30), followed by CellPLM (ARI = 0.82, NMI = 0.80, SIL = 0.26). Both models formed compact UMAP domains corresponding to the major aortic wall compartments. scELMo and Nicheformer also recovered the main cell populations, although with greater mixing at boundaries, whereas GenePT-w showed weaker separation (ARI = 0.50, SIL = 0.10). These results indicate that broad tissue cell types can be recovered in zero-shot mode when their expression programmes are well separated, but that gene-level semantic embeddings alone are less effective than expression-trained cellular representations.

The HS and BMMC datasets exposed more difficult regimes. HS contains closely related dendritic-cell states rather than broad lineages, and all models showed reduced performance compared with ATAA. scGPT-spatial remained strongest (ARI = 0.47, NMI = 0.56), followed by CellPLM (ARI = 0.41, NMI = 0.52), whereas Nicheformer and GenePT-w performed less well (ARI = 0.21 for both). scELMo produced the highest silhouette score (0.27), suggesting compact local structure, but its lower ARI and NMI indicate that compactness did not fully align with curated immunological labels. BMMC, comprising 45 haematopoietic populations across donors, sites and technologies, further shifted the ranking. scELMo achieved the highest ARI and silhouette score (ARI = 0.50, SIL = 0.19), while CellPLM produced the highest NMI (0.69), followed by Nicheformer (0.64) and scGPT-spatial (0.63). This metric-dependent pattern suggests that several models captured broad haematopoietic lineage structure, but differed in how they split or merged rare, transitional and batch-confounded populations. GenePT-w was consistently least robust in this setting (ARI = 0.12, SIL = −0.26).

Spatial transcriptomics posed a different challenge because successful clustering requires recovery of tissue architecture, not only transcriptional similarity. In HBCA1 breast cancer, expert annotations distinguish invasive tumour, tumour edge, DCIS/LCIS and adjacent healthy regions. scELMo achieved the highest ARI (0.35), whereas Novae obtained the highest NMI and silhouette score (NMI = 0.46, SIL = 0.19). Spatial maps showed that Novae and scELMo generated relatively contiguous domains that broadly followed tumour and peri-tumour regions, while scGPT-spatial and Nicheformer captured tumour-associated structure with more fragmented boundaries. CellPLM produced separated but partly over-segmented regions, and GenePT-w showed substantial spatial mixing. These patterns suggest that tumour-domain recovery benefits from models that encode spatial neighbourhoods or richer biological context, but that zero-shot embeddings still struggle with histological transition zones such as tumour edges.

In the MHPC hippocampus, quantitative agreement with atlas labels was modest for all models, with CellPLM showing the highest ARI and NMI (ARI = 0.13, NMI = 0.20) and scELMo showing the highest silhouette score (SIL = 0.30). However, the spatial plots revealed that several models, including CellPLM, scGPT-spatial, scELMo and Novae, preserved stripe-like organisation consistent with hippocampal lamination, whereas GenePT-w produced diffuse, poorly organised domains. This result highlights an important limitation of conventional clustering metrics in spatial benchmarks: discrete atlas labels may not fully capture continuous cytoarchitectural gradients, so anatomically plausible laminar structure can be penalised when it does not align exactly with coarse annotations.

In the MOSTA E9.5 embryo section, the benchmark tested recovery of organ-scale developmental compartments, including heart, liver, notochord, mesenchyme and brain. CellPLM achieved the highest ARI (0.26), while CellPLM and Novae shared the highest NMI (0.39), indicating the strongest concordance with annotated embryonic regions. Their spatial maps better preserved organ-scale structure, including central cardiac regions, axial domains and surrounding mesenchyme. Nicheformer and scGPT-spatial recovered several anatomical regions but with less distinct boundaries. scELMo achieved the highest silhouette score (0.31) but lower ARI and NMI, suggesting that it captured compact developmental gradients rather than the discrete organ labels used as ground truth. GenePT-w again failed to recover coherent anatomical compartments.

Together, these zero-shot analyses show that pretrained foundation model embeddings can recover meaningful biological structure, but the type of structure recovered depends strongly on modality, tissue complexity and model architecture. Expression-trained cell-level transformers, particularly scGPT-spatial and CellPLM, were most effective for resolving transcriptional cell identities in scRNA-seq datasets, whereas spatial or graph-aware models such as Novae, together with CellPLM and scELMo in selected settings, better preserved tissue-scale organisation. The decline in performance from broad vascular cell types to dendritic-cell subtypes, multi-site bone marrow populations and spatial tissue domains shows that zero-shot generalisation remains conditional rather than universal. This establishes a central benchmark principle: foundation models should be evaluated across multiple biological regimes and modalities, because strong performance in one setting does not guarantee transferable biological representation in another.

**Continual pretraining provides selective gains rather than universal improvement**

We next tested whether short domain-specific continual pretraining improves the biological structure encoded by foundation model embeddings. Each applicable model was adapted to each dataset for two self-supervised epochs using its default pretraining objective, after which embeddings were recomputed and evaluated with the same Leiden clustering workflow used in the zero-shot analysis. This comparison was restricted to models with directly applicable continual pretraining procedures for clustering: Nicheformer, CellPLM and scGPT-spatial for scRNA-seq datasets, and these models plus Novae for spatial transcriptomics datasets. Performance was assessed using ARI, NMI and silhouette score, enabling direct comparison with the zero-shot setting (**Figure 3a, 3b; Table 3**). In the single-cell datasets, continual pretraining produced strongly architecture- and dataset-dependent effects. In ATAA, where major aortic wall populations were already well separated in zero-shot embeddings, performance was largely saturated. scGPT-spatial retained the highest ARI and NMI (ARI = 0.87, NMI = 0.85), although its silhouette score decreased, indicating that label concordance was preserved but local cluster compactness was not improved. CellPLM showed only a small gain relative to zero-shot clustering (ARI 0.82 to 0.84; NMI 0.80 to 0.82), whereas Nicheformer decreased in ARI and NMI (ARI 0.62 to 0.57; NMI 0.78 to 0.74). These results suggest that when broad tissue compartments are already encoded in the pretrained space, additional self-supervised adaptation provides limited benefit and may even reduce alignment with curated labels for some architectures.

The strongest benefit of continual pretraining was observed in the HS dendritic-cell dataset, where labels correspond to subtle immunological states rather than broad lineages. scGPT-spatial improved markedly after adaptation, increasing from ARI 0.47 to 0.68, NMI 0.56 to 0.62 and SIL 0.15 to 0.34. The corresponding UMAPs showed clearer separation of closely related dendritic-cell subsets, consistent with refinement of transcriptional boundaries along the DC continuum. In contrast, CellPLM remained broadly stable but did not improve substantially, with ARI decreasing from 0.41 to 0.37 while NMI remained 0.52. Nicheformer declined further (ARI 0.21 to 0.14; NMI 0.34 to 0.23), suggesting that the same adaptation protocol can compress rather than resolve fine immune-state structure depending on the model architecture.

BMMC provided a more confounded test case because it contains 45 haematopoietic populations across donors, sites and technologies. Here, continual pretraining did not uniformly improve performance. scGPT-spatial showed small but consistent gains in NMI and silhouette score (NMI 0.63 to 0.66; SIL 0.14 to 0.19), suggesting slightly sharper separation of major haematopoietic lineages. CellPLM, however, decreased across all three metrics (ARI 0.31 to 0.28; NMI 0.69 to 0.66; SIL 0.13 to 0.12), and Nicheformer also declined in ARI and NMI. Thus, short self-supervised adaptation was insufficient to resolve the full complexity of rare, transitional and batch-confounded bone marrow populations, and in some cases reduced performance relative to the pretrained representation.

The spatial datasets showed a similarly selective pattern. In HBCA1 breast cancer, Novae benefited most from continual pretraining, improving from ARI 0.25 to 0.31 and NMI 0.46 to 0.51, while maintaining coherent tumour-associated domains. By contrast, Nicheformer and CellPLM changed only modestly, and scGPT-spatial decreased in label concordance (ARI 0.28 to 0.17; NMI 0.45 to 0.33). This suggests that adaptation can strengthen spatial-domain structure when the model objective is closely matched to tissue-neighbourhood organisation, but may destabilise histological boundary recovery in architectures less specialised for graph-like tissue segmentation. In the MHPC hippocampus, quantitative agreement with atlas labels remained low after continual pretraining, reflecting the difficulty of mapping continuous laminar cytoarchitecture onto discrete labels. CellPLM showed a small improvement in ARI and NMI (ARI 0.13 to 0.15; NMI 0.20 to 0.21), while scGPT-spatial improved in NMI and silhouette score despite little change in ARI. Novae retained similar ARI but showed a strong increase in silhouette score (0.15 to 0.32), indicating more compact spatial domains. These results suggest that continual pretraining can improve local spatial coherence or laminar continuity without necessarily increasing agreement with coarse anatomical labels. In the MOSTA E9.5 embryo, CellPLM remained the strongest model by ARI and NMI after continual pretraining (ARI = 0.27, NMI = 0.40), consistent with preservation of organ-scale developmental compartments. scGPT-spatial also improved relative to zero-shot clustering (ARI 0.19 to 0.21; NMI 0.30 to 0.36; SIL 0.08 to 0.17), suggesting that domain adaptation helped refine embryonic spatial structure. Novae showed lower ARI and NMI after adaptation but a higher silhouette score, indicating that it produced more compact domains that did not align as closely with the discrete organ annotations. This divergence again shows that self-supervised adaptation may optimise spatial coherence or developmental gradients rather than direct label concordance.

Together, these results show that continual pretraining is not a default route to better foundation model performance. Its benefit depends on the match between model architecture, pretraining objective and the biological structure of the target dataset. scGPT-spatial gained most clearly in fine-grained immune subtyping and selected complex settings, CellPLM was comparatively stable and remained strong in broad tissue and embryo analyses, Novae benefited in tumour-domain recovery, and Nicheformer often degraded under the short adaptation protocol. This establishes an important benchmark message: future foundation models should not only support continual pretraining, but should be designed and evaluated for when adaptation improves biological resolution, when it merely increases spatial compactness, and when it risks overfitting or eroding transferable structure.

## Supervised annotation reveals fine-grained and task-specific failure modes

Supervised cell type annotation evaluates a different aspect of foundation model utility from zero-shot clustering: whether pretrained representations can be adapted to expert-defined cellular taxonomies when labels are available. Because supervised performance depends on label availability, classifier design and each model's recommended fine-tuning workflow, we interpret this task as a measure of practical supervised adaptation utility, rather than intrinsic zero-shot generalisation. We fine-tuned or adapted each model on ground-truth labels from three scRNA-seq datasets, ATAA, HS and BMMC, and one single-cell-resolution spatial MERFISH atlas, SEAAD. Stratified splits were used to preserve a shared held-out test set across models, with a validation split where required. Performance was summarised using accuracy, precision, recall, macro F1, PR-AUC and ROC-AUC (**Table 4**). Because global accuracy and ROC-AUC can obscure failures in imbalanced or fine-grained taxonomies, we also examined row-normalised confusion matrices, where diagonal structure indicates high per-class fidelity and off-diagonal mass highlights systematic misclassification (**Figure 4**; **Supplementary Figure S1**).

In ATAA, supervised annotation was nearly saturated, reflecting the strong transcriptional separation among major aortic wall populations. Nicheformer, CellPLM and scGPT-spatial each achieved 0.97 accuracy, with macro F1 scores of 0.94–0.95 and PR-AUC values of 0.98. scELMo and GenePT-w were slightly lower but still performed strongly, with accuracies of 0.95 and 0.92 and macro F1 scores of 0.89 and 0.85, respectively. The confusion matrices showed almost complete diagonal recovery of smooth muscle cells, fibroblasts, endothelial cells and major immune compartments, with only limited leakage between related stromal or immune states. Thus, once labels are available, most models can learn broad vascular cell identities, making ATAA more useful as a saturation control than as a stringent discriminator of supervised model performance.

The HS dendritic-cell dataset provided a more informative test because its seven labels represent closely related immune states rather than broad lineages. CellPLM achieved the strongest performance, with 0.93 accuracy, 0.88 macro F1 and 0.96 PR-AUC, followed by scGPT-spatial with 0.91 accuracy and 0.84 macro F1. Their confusion matrices showed the cleanest diagonal structure, indicating better separation of cDC1, cDC2, AS-DC and mitotic DC populations (**Figure 4**). By contrast, Nicheformer retained reasonable accuracy (0.83) but had a much lower macro F1 score (0.43), indicating uneven class-wise recovery despite correct assignment of abundant states. scELMo and GenePT-w showed intermediate but noisier performance, with broader off-diagonal signal among cDC2-like, activation-associated and proliferative DC states. Biologically, this shows that supervised accuracy alone can overstate performance in fine immune taxonomies: the key challenge is whether a model preserves subtle maturation, activation and proliferation axes rather than collapsing them into dominant dendritic-cell classes.

BMMC represented the strongest test of rare-state sensitivity and class imbalance, with 45 annotated bone marrow populations spanning stem and progenitor cells, myeloid cells, lymphoid subsets, plasma cells and transitional states across donors, sites and technologies. CellPLM achieved the best balance across classes, with 0.86 accuracy and 0.73 macro F1. scGPT-spatial and Nicheformer reached similar accuracies of 0.84 but lower macro F1 scores of 0.65 and 0.61, respectively. Notably, scGPT-spatial achieved the highest PR-AUC in this dataset, suggesting strong ranking of class probabilities, but its lower macro F1 indicates that thresholded predictions still missed or merged some minority populations. Confusion matrices showed that most models recovered abundant lineages, including T cells, B cells, NK cells, monocytes and neutrophils, but differed in their handling of rare progenitors, plasmablasts and closely related memory subsets. GenePT-w and scELMo showed greater off-diagonal structure, consistent with absorption of rare or transitional states into broader haematopoietic categories. These results indicate that supervised fine-tuning improves coarse lineage annotation, but only selected models maintain sensitivity to rare and developmentally related populations.

We next evaluated SEAAD to test supervised annotation in a spatial, single-cell-resolution brain atlas. This dataset contains neuronal and non-neuronal cell types from the middle temporal gyrus, providing a distinct challenge from dissociated scRNA-seq because cell identity is embedded within cortical and disease-associated spatial context. Nicheformer and scGPT-spatial achieved the strongest global performance, both reaching 0.94 accuracy, with macro F1 scores of 0.88–0.89 and PR-AUC values of 0.95–0.97. Their confusion matrices showed strong recovery of excitatory and inhibitory neuronal classes, oligodendrocytes, OPCs, astrocytes, microglia and vascular cells, with residual errors concentrated among biologically related neuronal or glial subclasses. CellPLM remained competitive, with 0.91 accuracy and 0.84 macro F1, whereas GenePT-w and scELMo showed lower class-wise performance despite reasonable global accuracy. Novae was included in this task to test whether a spatial-domain representation can be repurposed for per-cell taxonomy; however, it performed poorly on SEAAD annotation, with 0.30 accuracy, 0.03 macro F1 and ROC-AUC of 0.50. This result should not be interpreted simply as model failure, but as evidence of a mismatch between representation scale and task: embeddings optimised for mesoscale tissue-domain discovery do not necessarily encode the discriminative features required for cell type classification.

This analysis shows that foundation models can achieve high performance when expert labels are available, but the apparent success depends strongly on biological granularity and class balance. Broad cell identities in ATAA were readily learned by most models, whereas HS, BMMC and SEAAD exposed differences in fine immune-state resolution, rare haematopoietic population recovery and spatial brain cell taxonomy. CellPLM and scGPT-spatial were the most consistently strong across single-cell annotation tasks, Nicheformer was particularly competitive in the SEAAD spatial brain atlas, and Novae highlighted the distinction between spatial-domain representation and per-cell classification. This establishes an important benchmark principle: supervised performance depends not only on the presence of labels, but also on whether the model's pretraining objective and representation scale match the biological resolution of the annotation task.

### Marker-gene concordance reveals biological specificity of spatial clusters

Clustering metrics quantify agreement with annotated domains, but they do not show whether model-derived clusters are biologically interpretable. We therefore asked whether clusters obtained after continual pretraining preserved the marker-gene programmes of annotated tissue domains. For each model, we identified the top 20 differentially expressed genes for each Leiden cluster and each ground-truth domain using Wilcoxon rank-sum tests, then compared these marker sets using raw overlap counts and Jaccard indices. We visualised the raw overlap between model-derived and annotation-derived marker sets in the MOSTA E9.5 embryo section as heatmaps in **Figure 5**, with corresponding normalised Jaccard comparisons shown in **Supplementary Figure S2**. This analysis provides an orthogonal test of biological validity: a useful embedding should not only assign cells or spots to the correct region, but should also recover the transcriptional programmes that define that region.

MOSTA E9.5 provides a stringent developmental benchmark because organ primordia such as brain, heart, liver, notochord, dermomyotome, sclerotome and mesenchyme are spatially organised but transcriptionally interrelated. Across models, the clearest marker concordance was observed for large, strongly patterned organs, particularly brain and heart (**Figure 5**). CellPLM, scGPT-spatial and Novae each produced clusters whose top markers overlapped extensively with the ground-truth brain or heart signatures, often sharing 13–18 of the top 20 genes. These strong diagonal-like signals indicate that the corresponding clusters captured organ-specific transcriptional programmes rather than only spatially contiguous regions.

The Jaccard analysis further refined this interpretation by normalising marker overlap by gene-set size (**Supplementary Figure S2**). For Nicheformer, CellPLM and scGPT-spatial, brain and heart showed the strongest normalised concordance, indicating that these clusters were relatively specific to the correct developmental programme. Novae showed particularly high specificity for heart, consistent with its graph-based spatial objective and its ability to preserve compact organ-scale domains. In contrast, GenePT-w and scELMo showed more diffuse overlap patterns, suggesting that although they recovered some organ-associated genes, their clusters mixed markers from neighbouring embryonic structures and were therefore less biologically specific.

Domains with shared developmental origin showed broader off-diagonal structure in the heatmaps (**Figure 5**). Mesenchyme, dermomyotome, sclerotome, branchial arch and related axial tissues did not map as cleanly to single clusters as brain or heart. Instead, their marker sets overlapped with several predicted clusters, reflecting the biological continuum of mesodermal differentiation and shared extracellular-matrix, migration and patterning programmes across these regions. This off-diagonal signal should therefore not be interpreted only as model error; it also reflects real developmental relatedness between adjacent primordia. Nevertheless, CellPLM and Novae produced the most specific separation of mesenchymal and structured mesoderm-derived domains, whereas more diffuse overlap patterns indicated weaker resolution of these closely related developmental states.

Small or heterogeneous regions, including AGM and liver, showed lower marker concordance across models (**Figure 5**). This likely reflects a combination of limited spot numbers, transitional cellular composition and partially shared haemato-endothelial or mesenchymal gene programmes. These observations highlight an important limitation of DEG-based benchmarking: marker concordance depends not only on embedding quality, but also on tissue size, annotation granularity and whether the annotated region corresponds to a discrete lineage programme or a developmental transition.

Overall, the DEG-based analysis shows that the strongest embeddings preserve biologically meaningful developmental programmes rather than merely matching spatial labels. Cell-level and spatial graph-based models produced the clearest organ- and lineage-focused signatures, whereas weaker or more diffuse overlap patterns revealed loss of marker specificity even when clusters appeared visually plausible. This analysis therefore adds an essential biological layer to the benchmark: foundation models should be evaluated not only by ARI, NMI or silhouette score, but also by whether their clusters retain interpretable gene programmes that reflect tissue identity, developmental origin and spatial organisation.

## Perturbation prediction separates expression reconstruction from functional response modelling

Perturbation prediction provides a functional stress test of foundation models by asking whether learned representations can predict transcriptional responses to genetic perturbation, rather than only organise static cell states. Following the scGPT-style perturbation evaluation framework[9], we evaluated CellPLM, scGPT-spatial, GenePT-w and scELMo on three Perturb-seq benchmarks: Adamson, Norman and Replogle[21-23]. These datasets test distinct perturbational regimes in K562 cells, including unfolded-protein-response perturbations, single- and combinatorial-gene perturbations and genome-scale CRISPRi. Nicheformer and Novae were not included in this task because they do not provide directly comparable perturbation-prediction workflows for this GEARS-based setting. To ensure direct comparability, all models were evaluated on the same gene set within each dataset, defined as the shared intersection between the Perturb-seq gene space and the gene vocabularies or embedding spaces of the evaluated models. For GenePT-w and scELMo, the exact model-derived gene embeddings were used as gene-feature inputs to a shared GEARS perturbation decoder, allowing us to test the perturbation-prediction utility of their embedding spaces under a harmonised downstream framework. Performance was assessed using MSE and Pearson correlation across all genes and top differentially expressed genes, together with delta-based metrics that measure perturbation-induced change relative to control (**Figure 6a, 6b**; **Table 5**).

Across all three datasets, global reconstruction metrics were high for every model, with Pearson correlations of approximately 0.98–0.99 and low MSE values across all genes (**Table 5**). This indicates that all evaluated models can approximate the overall post-perturbation expression profile. However, global expression reconstruction is not sufficient to establish biologically meaningful perturbation modelling, because most genes do not change substantially after perturbation. Response-focused metrics were therefore more informative. We interpreted Pearson Delta DE as the primary biological readout, supported by Pearson Delta and MSE DE, because these metrics test whether a model captures perturbation-induced changes in the genes most affected by the perturbation. In contrast, global MSE and Pearson correlation were treated as reconstruction-level sanity checks.

In the Norman dataset, model rankings depended strongly on the metric. scGPT-spatial achieved the lowest global MSE (0.0041), and scELMo achieved the highest global Pearson correlation (0.9880), indicating strong overall expression reconstruction. However, GenePT-w performed best on the response-focused metrics, with the lowest MSE DE (0.1757) and the highest Pearson DE (0.9252), Pearson Delta (0.5871) and Pearson Delta DE (0.7389). scELMo was the next strongest model by Pearson Delta DE (0.7032). These results suggest that, in a combinatorial perturbation setting, literature-derived gene embeddings can provide useful functional priors for predicting perturbation-responsive programmes, even when they are not the best global expression reconstructors.

The Adamson dataset showed a different pattern. GenePT-w achieved the lowest global MSE (0.0068), the highest global Pearson correlation (0.9896) and the highest Pearson DE (0.9576), indicating strong reconstruction of expression structure among response genes. However, scELMo achieved the lowest MSE DE (0.2104) and the highest Pearson Delta DE (0.7450), while CellPLM achieved the highest transcriptome-wide Pearson Delta (0.6487). scGPT-spatial remained competitive, particularly for Pearson DE (0.9369) and Pearson Delta DE (0.7097), but did not dominate the response-specific metrics. Thus, in this unfolded-protein-response-associated perturbation setting, language-derived gene-function embeddings were particularly effective for DE-gene response prediction, whereas CellPLM better captured the broader perturbation-induced shift across the transcriptome.

The Replogle genome-scale CRISPRi dataset was the most challenging setting for perturbation-response prediction. All models retained high global Pearson correlation, but delta-based metrics were substantially lower than in Norman and Adamson, indicating that genome-scale perturbations expose limitations in predicting effect sizes relative to control. CellPLM achieved the strongest delta-based performance, with the highest Pearson Delta (0.3801) and Pearson Delta DE (0.4604), closely followed by scELMo (Pearson Delta DE = 0.4579) and GenePT-w (0.4505). GenePT-w achieved the lowest MSE DE (0.1365), while scGPT-spatial achieved the highest Pearson DE (0.9437). This divergence suggests that scGPT-spatial captured expression structure among DE genes, whereas CellPLM more accurately modelled perturbation-induced changes relative to the control state. The modest absolute delta correlations across all models indicate that functional response prediction remains difficult, particularly for genome-scale perturbation landscapes.

Representative perturbation-level plots further illustrate the separation between expression reconstruction and response modelling (**Figure 6b**). Across selected perturbations, predicted expression distributions often followed the direction of change for highly responsive genes, but models differed in how well they reproduced the magnitude and variability of the response. This is especially important biologically because perturbation effects are defined not only by which genes respond, but also by the amplitude and coordination of those responses across regulatory programmes.

These results show that perturbation prediction defines a performance axis distinct from clustering and supervised annotation. Models that organise static cell states well are not necessarily the strongest predictors of perturbation-induced transcriptional change, and no single architecture dominated across all datasets and metrics. GenePT-w and scELMo were strong on response-gene metrics in Norman and Adamson when their exact language-derived gene embeddings were evaluated through the shared perturbation decoder, whereas CellPLM was more robust for delta-based prediction in the genome-scale Replogle screen. scGPT-spatial remained competitive but was not uniformly superior despite the task following an scGPT-style evaluation template. These findings highlight a key benchmark principle: accurate reconstruction of post-perturbation expression is not sufficient for biological perturbation modelling. Future single-cell foundation models will need objectives that more directly learn regulatory response structure, perturbation effect size and context-dependent gene–gene interactions.

## Discussion and future work directions

Foundation models are increasingly presented as general-purpose engines for single-cell and spatial omics, but our benchmark shows that "foundation" remains a conditional property rather than an intrinsic consequence of model scale or pretraining. Across zero-shot clustering, continual pretraining, supervised annotation, marker-gene concordance and perturbation prediction, model rankings changed with modality, tissue context, biological granularity and evaluation metric. Expression-trained cell-level transformers were strongest for many cell-identity tasks, spatial and graph-aware models better preserved tissue-scale organisation, and language-derived gene embeddings remained useful for selected perturbation-response metrics. These findings indicate that current models learn useful but partial biological representations: they encode aspects of cellular state, tissue architecture or gene function, but do not yet provide uniformly transferable representations across biological scales. This aligns with recent zero-shot evaluations showing that single-cell foundation models can underperform simpler baselines and remain vulnerable to dataset and platform shift[17].

A central implication is that single-cell and spatial foundation models should be evaluated as out-of-distribution biological systems, not only as pretrained encoders. In biology, domain shift is not an edge case; it is the problem itself. Donor, tissue, disease state, species, assay chemistry, sequencing depth, spatial resolution, gene panel and annotation granularity all define different biological domains. Our results illustrate this directly: models that performed well on broad aortic wall populations did not necessarily resolve subtle dendritic-cell subsets, batch-confounded bone marrow states or spatial tissue domains. Future benchmarks should therefore include explicit leave-domain-out tests, including leave-one-platform-out, leave-one-tissue-out, leave-one-disease-out, cross-species and cross-modality transfer. Without such tests, high performance may reflect proximity to the pretraining distribution rather than true biological generalisation.

A second lesson is that model architecture encodes a biological hypothesis. Gene tokens, ranked-gene sequences, cell-level embeddings, graph nodes, spatial neighbourhoods and text-derived gene vectors each privilege a different scale of biology. This explains why no model dominated all tasks: a representation optimised for per-cell identity is not necessarily optimal for tissue-domain segmentation, and a spatial-domain model does not automatically become a per-cell classifier. The field should therefore move beyond the assumption that a single generic transcriptomic foundation model can solve all biological questions. Instead, progress may require problem-specific foundation models that encode the mechanisms relevant to a defined biological or clinical task.

Immunotherapy provides a clear example. Predicting immune-checkpoint response requires more than generic cell-state representation: it requires modelling tumour-intrinsic programmes, antigen presentation, T-cell exhaustion, myeloid suppression, stromal exclusion, vascular niches and therapy-specific mechanisms. Compass, a recent pan-cancer immunotherapy model, illustrates this direction by using a concept-bottleneck transformer to map tumour transcriptomes into interpretable immune concepts, although it is currently a preprint and is based on bulk rather than single-cell/spatial data[24]. The broader design principle is important: clinically useful biological foundation models may need task-specific priors, interpretable intermediate states and disease-focused pretraining rather than generic representation learning alone.

Most current models are also still too RNA-centric. Even spatial models often encode spatially resolved RNA rather than integrating chromatin accessibility, protein abundance, receptor signalling, morphology, genotype, histology and clinical metadata. This matters because biological decisions made from one modality can be confidently wrong. A tumour may appear immune-inflamed by RNA expression but show spatial T-cell exclusion histologically; a transcript may be abundant without corresponding protein activity; and a perturbation model may reconstruct expression while missing regulatory effect size. Recent spatial models such as Nicheformer and Novae show that tissue context is becoming central, while compositional foundation-model perspectives argue for modular integration of RNA, chromatin, protein, imaging, spatial context and text[13,20,25].

Evaluation must also become more biological. ARI, NMI, F1 and ROC-AUC are useful, but they do not establish whether a model preserves interpretable gene programmes or predicts functional responses. Our DEG analysis showed that visually plausible clusters can differ in marker-gene specificity, while the perturbation benchmark showed that global expression reconstruction is much easier than predicting perturbation-induced changes among responsive genes. This concern is now central in perturbation modelling: Ahlmann-Eltze et al. showed that deep perturbation models do not yet consistently outperform simple baselines, and Systema demonstrated that standard control-referenced metrics can overestimate biological response prediction because of systematic variation[26,27]. Future benchmarks should therefore report shared gene sets, pretraining-data overlap, simple baselines, leave-domain-out splits, uncertainty intervals, marker-programme concordance, perturbation-specific metrics and representative failure cases.

Continual pretraining and fine-tuning should likewise be treated as biological adaptation problems, not automatic improvements. In our benchmark, domain-specific pretraining improved some subtle settings but degraded or failed to improve others. This suggests that future models need adaptation strategies that protect against catastrophic forgetting, overfitting to batch structure and collapse of rare states. Parameter-efficient fine-tuning, test-time adaptation, invariant representation learning, uncertainty calibration and OOD detection should become part of the biological foundation-model toolkit, especially when models are deployed on new tissues, diseases or technologies.

Finally, agentic AI may help convert foundation models from static encoders into decision-support systems. Agentic systems should not replace biological foundation models; rather, they can orchestrate them. A single-cell/spatial agent could select the appropriate model for a task, detect OOD samples, compare predictions across modalities, query literature and pathway databases, run simple baselines, assess uncertainty and propose validation experiments. Recent work on agentic biomedical systems, virtual laboratories and automated single-cell workflow standardisation suggests that this direction is becoming technically feasible[28-30]. For single-cell and spatial omics, the strongest future systems may combine specialised foundation models with agentic reasoning, multimodal evidence integration and human-in-the-loop validation.

In conclusion, our benchmark argues for a shift in how progress in single-cell and spatial foundation modelling is defined. Current models should not be judged by a single leaderboard score, model scale or the breadth of their pretraining corpus alone. Instead, they should be evaluated by whether they generalise across biological domains, preserve interpretable gene programmes, remain calibrated under uncertainty, integrate complementary modalities and support perturbation-grounded reasoning. This shift is essential because biological deployment rarely occurs in-distribution: new tissues, diseases, technologies, perturbations and clinical contexts all test whether a model has learned transferable biology rather than dataset-specific structure. The next generation of models should therefore move beyond generic RNA-centric encoders toward problem-aware, multimodal and causally informed systems that can reason across cellular identity, tissue architecture, molecular regulation and experimental intervention. In this view, foundation models become most valuable not as universal replacements for existing analysis tools, but as components of context-aware virtual cell systems capable of guiding biological discovery and translational decision-making[31,32].

## Method sections

To ensure that differences in model performance reflected model design and biological representation rather than inconsistent preprocessing or evaluation, we implemented a harmonised benchmarking workflow spanning dataset curation, preprocessing, model-specific implementation, downstream task design and metric computation. The framework was applied across scRNA-seq, spatial transcriptomics and Perturb-seq datasets, enabling evaluation of six foundation models across zero-shot clustering, continual pretraining, supervised annotation, marker-gene concordance and perturbation prediction. Wherever possible, models were run using released checkpoints, recommended preprocessing and default or author-recommended hyperparameters. When a model did not provide a native implementation for a given task, we implemented a task-matched workflow that preserved the model's architectural assumptions and documented the adaptation.

### Model inclusion criteria

Models were selected to represent major foundation-model design principles in single-cell and spatial omics while remaining comparable under a harmonised evaluation framework. Inclusion required that a model provide reusable gene-, cell- or tissue-level representations, have public or reproducible implementation resources, and be applicable to at least one major benchmark task, including clustering, continual pretraining, supervised annotation, marker-gene concordance or perturbation prediction. We therefore focused on Nicheformer, CellPLM, scGPT-spatial, GenePT-w, scELMo and Novae. Methods primarily designed for data simulation, generative augmentation, regulatory-network inference, natural-language reporting or perturbation-only prediction were not included as primary foundation models, although they may serve as important baselines or complementary tools in future benchmark extensions.

## Datasets

### Single-cell datasets collection

**ATAA Dataset.** The ATAA dataset comprises scRNA-seq profiles from ascending aortic tissue from 11 individuals, including eight patients with ascending thoracic aortic aneurysm and three controls. The original study identified 11 major cell types through integrative clustering and marker-based annotation, which were used as reference labels.

**BMMC Dataset.** The BMMC dataset was generated for the NeurIPS 2021 Multimodal Single-Cell Data Integration Challenge and contains bone marrow mononuclear cells from 12 healthy donors. We used the CITE-seq subset, which was profiled across multiple sites with a nested batch design. Curated marker-gene and expert annotations were used as reference labels, comprising 45 annotated cell types.

**HS Dataset.** The HS dataset comprised human splenic dendritic cells isolated from macroscopically normal spleen tissue obtained from patients undergoing tumour resection in non-splenic organs. Seven dendritic-cell subtypes were used as ground-truth labels for supervised annotation and clustering evaluation.

### Spatial transcriptomics datasets collection

All spatial transcriptomics datasets except SEAAD were obtained from the BenchmarkST data resource (https://benchmarkst-reproducibility.readthedocs.io/en/latest/Data%20availability).

**HBCA1 Dataset.** The Human Breast Cancer Block A Section 1 (HBCA1) dataset is a 10x Genomics Visium section of human breast cancer. We used the fine_annot_type annotations from Xu et al. (2024) as reference spatial-domain labels, comprising 20 domains: healthy, tumour edge, IDC and DCIS/LCIS subregions.

**MHPC Dataset.** The annotated Mouse Hippocampus (MHPC) dataset contains a single slice of the mouse hippocampus region sequenced using slide-seqV2 and is acquired from the Broad Institute. We use the 'cluster' annotations from the dataset file, which contains 14 domain labels: ca1 ca2 ca3 subiculum, dentate pyramids, astrocytes, interneurons, oligodendrocytes, subiculum entorhinal cl2, endothelial stalk, subiculum entorhinal cl3, polydendrocytes, endothelial tip, microglia, mural, neurogenesis and ependymal.

**MOSTA Embryo Dataset.** The Mouse Organogenesis Spatiotemporal Transcriptomic Atlas (MOSTA) profiles mouse organogenesis using Stereo-seq. We analysed the E9.5 E1S1 section, using 12 annotated anatomical regions, including brain, heart, liver, mesenchyme, notochord and AGM, as reference labels.

**SEAAD Dataset.** The Seattle Alzheimer's Disease Brain Cell Atlas (SEAAD) is a multimodal atlas of Alzheimer's disease and related dementias. We used the middle temporal gyrus MERFISH dataset, which includes samples from 27 aged donors, a 140-gene panel and 24 annotated cell types used as reference labels.

**Perturb-Seq datasets**

**Adamson Perturb-Seq dataset.** The Adamson dataset contains gene expression data of 87 single-gene perturbations from the K562 leukemia cell line perturbed by Perturb-seq[21].

**Norman Perturb-Seq dataset.** The Norman dataset contains gene expression data of 131 two-gene perturbations and 105 single-gene perturbations from the K562 leukemia cell line perturbed by Perturb-seq[22].

**Replogle Perturb-Seq dataset.** The Replogle dataset contains genome-wide perturbations in the K562 leukemia cell line with CRISPR interference[23].

**Downstream evaluation tasks.** We evaluated the models across five benchmarking tasks – zero-shot clustering, continually-pretrained clustering, supervised fine-tuning, biomarker gene comparison, and perturbation prediction – to assess unsupervised, self-supervised, and supervised performance on single-cell and spatial transcriptomics datasets.

**Ground truth visualisation.** Ground-truth UMAPs for scRNA-seq datasets were generated using Scanpy (v1.11.5). Cells were normalised to 10,000 counts per cell, log-transformed, filtered to retain genes expressed in at least three cells, and restricted to the top 2,000 highly variable genes using the cell_ranger method. Data were scaled, reduced by PCA, and visualised using UMAP based on a nearest-neighbour graph constructed from the top 40 principal components.

## Zero-shot cell type clustering

Zero-shot clustering was performed to evaluate whether pretrained model embeddings captured biologically meaningful structure without additional training on the target dataset. For each model–dataset pair, embeddings were extracted using the corresponding pretrained model and its recommended preprocessing procedure. Clustering was then performed on the resulting embedding space using Leiden community detection. The Leiden resolution was swept until the number of clusters matched the number of reference labels. For models that originally used an alternative clustering approach, such as k-means, we additionally replicated the reported method and included those results in the Supplementary Information (**Supplementary Table S2**). Clustering performance was assessed using adjusted Rand index (ARI), normalised mutual information (NMI) and silhouette score. For single-cell datasets, UMAPs of the model embeddings were generated for visualisation. For spatial transcriptomics datasets, spatial plots were generated using the provided tissue coordinates and model-derived cluster or domain assignments. Random seeds were fixed at 42 for NumPy, Scanpy and PyTorch before stochastic operations, including PCA and Leiden clustering.

**Nicheformer.** For Nicheformer, pretrained model weights were loaded and each dataset was stored as an AnnData object and concatenated with the model to ensure gene vocabulary alignment. Following the original model's protocols, a technology mean vector was calculated separately for each downstream dataset. Metadata tokens for modality, species, and assay were defined according to the provided dictionaries. Raw gene expression values for the entire dataset were passed through the model to generate cell embeddings, with dataset preprocessing following the model's internal preprocessing steps. Embeddings were generated using a batch size of 32.

**CellPLM.** For CellPLM, pretrained model weights were loaded and each dataset was stored as an AnnData object. Raw counts were normalised by library size and log1p-transformed prior to embedding generation. The resulting CellPLM embeddings were used for Leiden clustering and downstream evaluation.

**scGPT-spatial.** For scGPT-spatial, single-cell and spatial transcriptomics datasets were loaded as AnnData objects and embedded using the released pretrained checkpoint, including model weights and gene vocabulary. Genes were aligned to the checkpoint vocabulary, and out-of-vocabulary genes were discarded. Embeddings were generated using scgpt_spatial.tasks.embed_data in inference mode with batch size 64 and fast-transformer enabled. For each cell, the function constructed a sequence of non-zero genes and expression values, applied two-step normalisation consisting of global non-zero mean scaling followed by per-gene mean division, and used a data collator to pad sequences, bin expression values into 51 levels and prepend a CLS token. The transformer was run in evaluation mode using Flash-Attention, and the L2-normalised CLS hidden state was used as the cell embedding and stored in adata.obsm["X_scGPT"]. For single-cell datasets, Leiden clustering was performed directly on these embeddings. For spatial transcriptomics datasets, embeddings were first projected using PCA with 20 components, and the k-nearest-neighbour graph was constructed using n_neighbors = 50.

**GenePT-w.** For GenePT-w, datasets were loaded as AnnData objects and raw counts were normalised to 10,000 total counts per cell followed by log1p transformation. Gene-level representations were generated by embedding curated NCBI Gene summaries using GPT-3.5 text-embedding-ada-002, after removing hyperlinks and date information and augmenting summaries with HGNC aliases. This produced 1,536-dimensional gene vectors. Cell embeddings were computed by multiplying the normalised expression matrix by the corresponding gene embeddings and dividing by the number of expressed genes. A cosine k-nearest-neighbour graph was then constructed for Leiden clustering, and silhouette scores were computed using cosine distance. For completeness, k-means clustering with the number of clusters set to the number of reference labels was also evaluated and reported in the Supplementary Information.

**scELMo.** For scELMo, gene representations were generated by prompting GPT-3.5 to produce one-paragraph academic summaries of each gene's major function, including pathway information, followed by text embedding to obtain gene-level vectors. Cell embeddings were computed using the weighted-average mode, in which each cell's total expression count was scaled to 1 without log normalisation, preserving the relative contribution of expression values. These embeddings were then used for Leiden clustering and evaluation.

**Novae.** Novae was evaluated on spatial transcriptomics datasets only. Datasets were loaded as AnnData objects with spatial coordinates, and cell neighbourhoods were computed using novae.spatial_neighbors. A pretrained Novae model was loaded from the Hugging Face Hub, and zero-shot representations were generated using model.compute_representations(zero_shot=True), where the graph attention network encodes each cell's local microenvironment without task-specific retraining. Spatial domain assignments were obtained using model.assign_domains, which performs batch-effect correction and constructs a nested hierarchy of spatial domains without external tools such as Harmony or mclust. The number of assigned domains was set to match the number of reference labels.

For all models, a random seed of 42 was set for NumPy, Scanpy, and PyTorch prior to any stochastic operations, including PCA and Leiden clustering, to ensure reproducibility across runs.

### Continually-pretrained cell type clustering

Continual pretraining was used to assess whether short, self-supervised adaptation to each target dataset improved the biological structure of model embeddings. For each specific model–dataset pair, the model was continually pretrained in a self-supervised manner for two epochs on the corresponding dataset. All other hyperparameters were kept at their default or recommended fine-tuning settings as reported in the original studies. After continual pretraining, embeddings were extracted from the updated model and clustering was performed following the same protocols as those used for the zero-shot clustering experiments. The evaluation procedures were also identical to those used for zero-shot clustering, ensuring consistent comparison across tasks.

**Nicheformer.** For Nicheformer, pretrained weights were loaded and each dataset was stored as an AnnData object. To ensure gene vocabulary alignment, each dataset was concatenated with the provided model AnnData object following the original workflow. A technology mean vector was calculated separately for each downstream dataset, and metadata tokens for modality, species and assay were assigned using the provided dictionaries. The model was continually pretrained for two epochs using the default hyperparameters specified in the original paper and codebase. Cell embeddings were then extracted from the updated model and evaluated using the same clustering protocol as in the zero-shot setting.

**CellPLM.** For CellPLM, continual pretraining was implemented by fine-tuning the pretrained checkpoint (20231027_85M) on each target dataset. Because the original repository did not provide an explicit continual-pretraining pipeline, our implementation followed the model architecture and reconstruction objective described in the original publication. Raw counts were library-size normalised and log-transformed. Gene symbols were mapped to Ensembl identifiers and aligned to the pretrained CellPLM gene vocabulary. Genes absent from the dataset were added as zero-expression columns, and the dataset was reordered to match the pretrained gene order. The model was fine-tuned end-to-end for two epochs under the reconstruction objective using the Adam optimiser with a learning rate of $1\times10^{-4}$, weight decay of $1\times10^{-6}$, ReduceLROnPlateau scheduling and gradient norm clipping. Cell embeddings were then extracted from the fine-tuned model and evaluated using the shared clustering workflow.

**scGPT-spatial.** For scGPT-spatial, the released pretrained checkpoint was continually pretrained on each target dataset using the masked value completion objective. Raw counts were library-size normalised to 10,000 counts per cell, log1p-transformed and discretised into 51 expression bins, with gene tokens aligned to the checkpoint vocabulary. Training was performed for two epochs using the Adam optimiser with a learning rate of $1\times10^{-4}$ and a masking ratio of 0.20. The learning rate followed the authors' fine-tuning tutorial, and the two-epoch limit was chosen to balance domain adaptation against overfitting. The best checkpoint was selected based on validation masked-token mean squared error. Cell embeddings were recomputed using the updated weights and evaluated using the shared clustering workflow. For this model, the k-nearest-neighbour graph was constructed with $k = 15$.

**Novae.** Novae was continually pretrained on spatial transcriptomics datasets only. Datasets were loaded as AnnData objects with spatial coordinates, and cell neighbourhoods were computed using novae.spatial_neighbors. A pretrained Novae model was loaded from the Hugging Face Hub and fine-tuned for two epochs with the self-supervised graph attention objective and default training hyperparameters. Following fine-tuning, cell representations were recomputed using model.compute_representations, and spatial domain assignments were obtained using model.assign_domains. The number of assigned domains was set to match the number of reference labels.

### Cell type annotation

For this task, we performed supervised fine-tuning of each model on the cell type classification objective using reference cell type labels. Since reference cell type labels were required, this task was restricted to datasets with cell-level annotations, including three scRNA-seq datasets and the single-cell-resolution SEAAD MERFISH spatial atlas. For dataset partitioning, we applied stratified sampling with a default 8:2 train/test split. For models that required a validation set (i.e. CellPLM), the training portion was further split into 7:1 train/validation, resulting in a 7:1:2 train/validation/test partition. For SEAAD, which contains 69 tissue sections, we perform a section-wise split to avoid data leakage between train and test sets. For SEAAD's validation set, we randomly select two tissue sections from the train set as the validation. This strategy ensured that the test set remained consistent across all models, allowing for a fair comparison. To ensure reproducibility, a random seed of 42 was applied to all splitting and stochastic processes. All models were fine-tuned using their default or recommended hyperparameters as reported in their original publications or official repositories. Model performance was evaluated using global Accuracy, Precision, Recall, macro F1-score, macro-PR-AUC, and macro-ROC-AUC. For k-nearest-neighbour classifiers, class probabilities were estimated as the fraction of neighbours assigned to each class, which were then used to compute macro-PR-AUC and macro-ROC-AUC.

**Nicheformer.** Because Nicheformer did not evaluate cell type annotation in its original publication, we implemented this task by adapting the model source code and tutorial workflow. Single-cell datasets were loaded as AnnData objects and, where necessary, gene symbols were converted to Ensembl gene identifiers. Converted gene lists were made unique, and datasets were subset to genes shared with the pretrained Nicheformer vocabulary. To ensure compatibility with the expected feature space, each dataset was concatenated with the provided model AnnData object, after which the placeholder observation was removed. Required metadata fields for modality, assay and species were then assigned.

Cells without labels were excluded, and remaining cells were split into stratified training and test sets using the target cell-type labels. Labels were encoded as integer class codes and stored in adata.obs together with the split indicator. NicheformerDataset objects were constructed using the precomputed technology mean vector and the model tokenisation settings: max_seq_len = 1500, aux_tokens = 30 and chunk_size = 1000. Data were loaded using PyTorch dataloaders, with shuffled training batches and deterministic test batches, using a batch size of 9.

The pretrained Nicheformer backbone was fine-tuned for supervised classification using NicheformerFineTune, with a mean-pooling head applied to features extracted from transformer layer 11. Fine-tuning was performed for one epoch in PyTorch Lightning with a learning rate of $1\times10^{-4}$, one warm-up epoch, gradient norm clipping with maximum norm 1.0 and gradient accumulation over 10 steps. Backbone parameters were updated during training (freeze = False). All sources of randomness were fixed using a global seed of 42 for Python, NumPy, PyTorch and PyTorch Lightning, with deterministic CuDNN settings. Evaluation was performed on the held-out test set using a consistent category-to-index mapping between training and evaluation.

**CellPLM.** For CellPLM, we followed the cell type annotation tutorial provided in the official repository. Datasets were loaded as AnnData objects, normalised to 10,000 total counts per cell and log1p-transformed. Cells were split into training, validation and test sets using a stratified 70:10:20 partition, generated by first applying an 80:20 train/test split and then dividing the training portion into train and validation subsets. Pretrained model weights provided by the authors were loaded before fitting, and the recommended default hyperparameters were used throughout. Performance was evaluated on the held-out test set. UMAPs were also generated from the test-set gene expression data, with separate plots coloured by predicted and reference labels.

**scGPT-spatial.** For scGPT-spatial, we followed the authors' cell type annotation tutorial end to end, using the default variables and hyperparameters with the cell-type classification objective enabled. Because the input expression matrices contained raw UMI counts, data_is_raw was set to True. Cell-type label columns were harmonised across datasets before training. The model was trained for 10 epochs with a fixed random seed of 42. At test time, predicted labels were obtained by taking the argmax of the classifier output and mapping predicted and reference label indices back to cell-type names using the provided taxonomy. To compute AUC metrics, an additional inference pass was performed on the test set to extract classifier logits, which were converted to class probabilities using softmax.

**GenePT-w.** For GenePT-w, cell type annotation was implemented using a k-nearest-neighbour classifier. For each cell, a representation was constructed by computing the expression-weighted average of GPT-3.5 semantic gene embeddings where each gene's embedding vector was scaled by its expression level in that cell and summed across all genes shared between the dataset and the embedding vocabulary. Cells were then split into stratified training and test subsets using an 80:20 ratio. A k-nearest-neighbour classifier was implemented with hnswlib using cosine similarity and trained on the expression profiles of the training cells. For each test cell, nearest neighbours were identified, and the predicted label was assigned by majority vote over the neighbouring training-cell annotations.

**scELMo.** For scELMo, cell type annotation was performed by fine-tuning a task-specific adapter that combines GPT-3.5-derived gene embeddings with single-cell expression profiles. The adapter was trained using a composite loss consisting of cross-entropy classification loss and contrastive loss, with the contrastive term designed to separate cell representations across conditions in latent space. The contrastive loss was weighted by $\lambda = 100$ to prioritise label-aware clustering. Cells were split into training, validation and test sets using a 70:10:20 ratio. The adapter was trained for 500 epochs. After fine-tuning, a k-nearest-neighbour classifier was trained on the resulting training-set cell embeddings and used to predict labels for the held-out test cells.

**Novae.** Although Novae did not include cell type annotation as one of its original downstream applications, we included this task to enable consistent comparison across models. Pretrained Novae representations were extracted from the frozen encoder and used as fixed input features for a lightweight multilayer perceptron (MLP) classifier trained to predict cell-type labels. The MLP head was optimised using the Adam optimiser for one epoch, while the Novae encoder weights remained frozen throughout training. The dataset was partitioned into training and test sets using a stratified 80:20 split, and model performance was evaluated on the held-out test set.

### DEG Analysis

To assess the biological relevance of clustering results, we compared the biomarker genes identified from the continually pretrained clustering task with those from the ground-truth annotations. Using the Scanpy package, we performed Wilcoxon rank-sum tests on the Leiden clusters obtained from the continually pretrained embeddings to identify the top 20 differentially expressed genes (DEGs) per cluster, which were considered the biomarker genes for those clusters. The same procedure was repeated for the ground-truth cell-type labels, yielding 20 biomarker genes per cell type. We then computed pairwise Jaccard indices between the biomarker gene sets of the Leiden clusters and the ground-truth cell types to quantify their overlap.

### Perturbation Prediction

The perturbation prediction task evaluated whether foundation model-derived representations could support prediction of post-perturbation transcriptional states from perturbation identity and cellular expression context. We followed a scGPT-style perturbation evaluation framework[9] using three Perturb-seq datasets: Adamson, Norman and Replogle[21-23]. These datasets comprise genetic perturbation screens in K562 cells and provide a controlled benchmark for evaluating response prediction across different perturbation designs, including single-gene perturbations, combinatorial perturbations and genome-scale CRISPR interference.

This task was restricted to models with compatible perturbation-prediction workflows or adaptable gene-level embedding outputs: CellPLM, scGPT-spatial, GenePT-w and scELMo.

Nicheformer and Novae were not included because they do not provide directly comparable perturbation-prediction workflows for this GEARS-based setting. Perturb-seq datasets were loaded using the PertData interface from GEARS and partitioned using the standard simulation split with a fixed random seed of 42, ensuring that perturbation conditions in the test set were not observed during training. Training, validation and testing were performed using GEARS dataloaders, with model-specific training procedures described below.

To ensure direct comparability across models, all perturbation-prediction metrics were computed on the same evaluated gene set within each dataset. For each Perturb-seq dataset, this gene set was defined as the shared intersection between the dataset gene space and the gene vocabularies or embedding spaces of all evaluated models. The same gene set was therefore used for CellPLM, scGPT-spatial, GenePT-w and scELMo within each dataset, ensuring that differences in MSE, Pearson correlation and delta-based metrics were not driven by unequal gene coverage.

Model predictions were evaluated using the unified GEARS evaluation framework. For each held-out perturbation condition, predicted and observed post-perturbation expression profiles were compared across the shared evaluated gene set. Performance was assessed using mean squared error (MSE) and Pearson correlation across all genes, as well as MSE and Pearson correlation restricted to the top 20 differentially expressed genes for each perturbation condition. To specifically evaluate perturbation-induced transcriptional change, we also computed delta-based metrics, in which predicted and observed expression changes were calculated relative to the corresponding control expression profile. These included Pearson Delta across all evaluated genes and Pearson Delta DE across the top 20 differentially expressed genes. Metrics were averaged across held-out perturbation conditions.

**CellPLM.** For CellPLM, the original repository did not provide a complete perturbation prediction implementation, so we implemented the task following the methodological description in the original publication. Perturb-seq datasets were loaded using the GEARS PertData interface and partitioned using the standard simulation split. CellPLM was initialised from the released pretrained checkpoint (20231027_85M), and a perturbation prediction head was attached with output dimensionality matching the model gene vocabulary. Genes in each Perturb-seq dataset were aligned to the pretrained CellPLM vocabulary using a fixed mapping between dataset genes and the ordered model gene list. Predictions were subsequently projected to the shared evaluated gene set for metric computation.

Perturbed inputs were constructed by marking perturbed genes using a sentinel perturbation value before remapping to the model feature space. Batch tensors were packed into the CellPLM input dictionary format, including sparse input and label matrices and an overlap-based input mask. The model was fine-tuned end-to-end to minimise MSE between predicted and observed post-perturbation expression values. Optimisation used AdamW, with a lower learning rate for the embedder than for the encoder, latent module and prediction head, together with ReduceLROnPlateau scheduling, weight decay, early stopping based on validation loss, linear learning-rate warm-up and global gradient norm clipping. Predictions and targets were projected back to the shared GEARS gene space for final evaluation.

**scGPT-spatial.** For scGPT-spatial, perturbation prediction was implemented following the authors' scGPT perturbation tutorial and aligned to the GEARS evaluation framework. Perturb-seq datasets were loaded using PertData and partitioned using the standard simulation split with the same fixed random seed. scGPT-spatial was initialised from the released pretrained checkpoint, including the encoder, value encoder and transformer encoder weights. Genes were aligned to the checkpoint vocabulary, and the model was fine-tuned using the masked gene expression reconstruction objective, in which perturbed genes were masked and the model was trained to reconstruct post-perturbation expression values.

Fine-tuning was performed using Adam optimisation with learning-rate scheduling, batch size 64 and 15 training epochs, following the scGPT perturbation workflow. Fast-Attention was enabled during training and inference. At evaluation, predicted and observed post-perturbation profiles were compared on the same shared evaluated gene set used for all models. Metrics were computed using the GEARS compute_metrics and deeper_analysis functions, including all-gene, DE-gene and delta-based measures.

**GenePT-w.** For GenePT-w, perturbation prediction was evaluated by using the exact GenePT-w gene embeddings as gene-feature inputs to a shared GEARS perturbation decoder. GenePT-w embeddings were derived from GPT-3.5 text embeddings of curated NCBI gene summaries, as described for the other GenePT-w analyses in this benchmark. Because GenePT-w does not provide a native end-to-end perturbation prediction decoder in the same format as GEARS or scGPT, this implementation evaluates the perturbation-prediction utility of the GenePT-w embedding space under a harmonised downstream decoder.

Perturb-seq datasets were loaded using the GEARS PertData interface and partitioned using the standard simulation split. The default GEARS gene features were replaced with the corresponding GenePT-w embeddings for genes in the shared evaluated gene set, while the remaining GEARS architecture, training objective and default hyperparameters were retained. The model was trained to predict post-perturbation expression profiles, and predictions were evaluated using the same GEARS metric functions as the other models. Thus, performance reflects the utility of GenePT-w-derived gene representations within a common perturbation-prediction framework.

**ScELMo.** For scELMo, perturbation prediction was evaluated using the authors' GEARS-based implementation to maintain consistency with the perturbation-prediction framework used across the benchmark. This implementation modifies the original GEARS architecture by incorporating LLM-derived gene embeddings into the gene representation layer while retaining the remaining GEARS perturbation-prediction pipeline. Specifically, pretrained scELMo gene embeddings derived from GPT-3.5-generated functional gene summaries were projected into the GEARS hidden space through a learnable adapter layer and combined with an additional learnable gene representation to preserve task-specific flexibility during training. The pretrained LLM embeddings remained fixed, while the adapter and downstream GEARS components were trained on the Perturb-seq data. All other GEARS components were kept unchanged. Evaluation was performed using the same shared evaluated gene sets and the same all-gene, DE-gene and delta-based metrics used for the other models.

## Author contributions

HAR and YY designed and conceptualized the study; SC, HAR, YY wrote the paper. HAR, SC, YY, FV, AA, NF, NL, MY, AB, RZ, TP, and MG edited the manuscript. SC, RZ, LC, MB, and RN carried out all the analyses and generated all the figures and tables with the supervision of HAR.

## Conflict of interest

The authors declare no competing financial and non-financial interests.

## Funding

This study was supported by the UNSW Scientia Program Fellowship and the Australian Research Council Discovery Early Career Researcher Award (DECRA), under grant

DE220101210 to H.A.R. S.C. was additionally supported by the CSIRO Next Generation Graduate Program (NGGP) in partnership with 23Strands. This collaborative project was also supported by the Second Century Fund (C2F), Chulalongkorn University. Additionally, the project has support from the Ratchadaphiseksomphot Endowment Fund, Top-up Grant of C2F, Chulalongkorn University.

## Data and Tool Availability

The code for preprocessing, running the models and evaluation can be accessed at https://github.com/iced-soda/FM-benchmark.

## References

1 Jovic, D. *et al.* Single-cell RNA sequencing technologies and applications: A brief overview. *Clinical and Translational Medicine* **12**, e694 (2022). https://doi.org/10.1002/ctm2.694

2 Hwang, B., Lee, J. H. & Bang, D. Single-cell RNA sequencing technologies and bioinformatics pipelines. *Exp Mol Med* **50**, 1–14 (2018). https://doi.org/10.1038/s12276-018-0071-8

3 Ståhl, P. L. *et al.* Visualization and analysis of gene expression in tissue sections by spatial transcriptomics. *Science* **353**, 78–82 (2016). https://doi.org/10.1126/science.aaf2403

4 Burgess, D. J. Spatial transcriptomics coming of age. *Nat Rev Genet* **20**, 317–317 (2019). https://doi.org/10.1038/s41576-019-0129-z

5 Cheng, M. *et al.* Spatially resolved transcriptomics: a comprehensive review of their technological advances, applications, and challenges. *Journal of Genetics and Genomics* **50**, 625–640 (2023). https://doi.org/10.1016/j.jgg.2023.03.011

6 Shi, W. *et al.* Next-Generation Sequencing-Based Spatial Transcriptomics: A Perspective from Barcoding Chemistry. *JACS Au* **4**, 1723–1743 (2024). https://doi.org/10.1021/jacsau.4c00118

7 Baek, S., Song, K. & Lee, I. Single-cell foundation models: bringing artificial intelligence into cell biology. *Exp Mol Med* **57**, 2169–2181 (2025). https://doi.org/10.1038/s12276-025-01547-5

8 Qiu, P. *et al.* BioLLM: A standardized framework for integrating and benchmarking single-cell foundation models. *Patterns* **6** (2025/08/08). https://doi.org/10.1016/j.patter.2025.101326

9 Cui, H. *et al.* scGPT: toward building a foundation model for single-cell multi-omics using generative AI. *Nature Methods 2024 21:8* **21** (2024–02–26). https://doi.org/10.1038/s41592-024-02201-0

10 Theodoris, C. V. *et al.* Transfer learning enables predictions in network biology. *Nature* **618**, 616–624 (2023). https://doi.org/10.1038/s41586-023-06139-9

11 Chen, Y. & Zou, J. Simple and effective embedding model for single-cell biology built from ChatGPT. (2025). https://doi.org/10.1038/s41551-024-01284-6

12 Wen, H. *et al.* in *International Conference on Learning Representations* Vol. 2024 5649–5673 (2024).

13 Tejada-Lapuerta, A. *et al.* Nicheformer: a foundation model for single-cell and spatial omics. *Nat Methods* **22**, 2525–2538 (2025). https://doi.org/10.1038/s41592-025-02814-z

14 Bian, H. *et al.* General-purpose pre-trained large cellular models for single-cell transcriptomics. *National Science Review* **11**, nwae340 (2024). https://doi.org/10.1093/nsr/nwae340

15 Csendes, G., Sanz, G., Szalay, K. Z. & Szalai, B. Benchmarking foundation cell models for post-perturbation RNA-seq prediction. *BMC Genomics* **26**, 393 (2025). https://doi.org/10.1186/s12864-025-11600-2
16 Wu, J. *et al.* Biology-driven insights into the power of single-cell foundation models. *Genome Biology* **26**, 334 (2025). https://doi.org/10.1186/s13059-025-03781-6
17 Kedzierska, K. Z., Crawford, L., Amini, A. P. & Lu, A. X. Zero-shot evaluation reveals limitations of single-cell foundation models. *Genome Biology* **26**, 101 (2025). https://doi.org/10.1186/s13059-025-03574-x
18 Liu, T. *et al.* Embeddings from language models are good learners for single-cell data analysis. *Patterns* **7** (2026/02/13). https://doi.org/10.1016/j.patter.2025.101431
19 Wang, C. *et al.* scGPT-spatial: Continual Pretraining of Single-Cell Foundation Model for Spatial Transcriptomics. *bioRxiv* (2025–02–08). https://doi.org/10.1101/2025.02.05.636714
20 Blampey, Q. *et al.* Novae: a graph-based foundation model for spatial transcriptomics data. *Nature Methods 2025 22:12* **22** (2025–12–10). https://doi.org/10.1038/s41592-025-02899-6
21 Adamson, B. *et al.* A Multiplexed Single-Cell CRISPR Screening Platform Enables Systematic Dissection of the Unfolded Protein Response. *Cell* **167**, 1867–1882.e1821 (2016). https://doi.org/10.1016/j.cell.2016.11.048
22 Norman, T. M. *et al.* Exploring genetic interaction manifolds constructed from rich single-cell phenotypes. *Science* **365**, 786–793 (2019). https://doi.org/10.1126/science.aax4438
23 Replogle, J. M. *et al.* Mapping information-rich genotype-phenotype landscapes with genome-scale Perturb-seq. *Cell* **185**, 2559–2575.e2528 (2022). https://doi.org/10.1016/j.cell.2022.05.013
24 Shen, W. *et al.* Generalizable AI predicts immunotherapy outcomes across cancers and treatments. *medRxiv* (2025 May 5). https://doi.org/10.1101/2025.05.01.25326820
25 Bahrami, M., Richter, T., Schmacke, N. A., Lavandera, A. E. & Theis, F. J. From modality-specific to compositional foundation models for cell biology. *Cell Systems* **17** (2026/02/18). https://doi.org/10.1016/j.cels.2026.101534
26 Ahlmann-Eltze, C. *et al.* Deep-learning-based gene perturbation effect prediction does not yet outperform simple linear baselines. *Nature Methods 2025 22:8* **22** (2025–08–04). https://doi.org/10.1038/s41592-025-02772-6
27 Viñas Torné, R. *et al.* Systema: a framework for evaluating genetic perturbation response prediction beyond systematic variation. *Nature Biotechnology 2025* (2025–08–25). https://doi.org/10.1038/s41587-025-02777-8
28 Swanson, K. *et al.* The Virtual Lab of AI agents designs new SARS-CoV-2 nanobodies. *Nature 2025 646:8085* **646** (2025–07–29). https://doi.org/10.1038/s41586-025-09442-9
29 Li, B. *et al.* Agentic AI and the rise of in silico team science in biomedical research. *Nature Biotechnology 2026* (2026–02–24). https://doi.org/10.1038/s41587-026-03035-1
30 Nouri, N. *et al.* An agentic AI framework for ingestion and standardization of single-cell RNA-seq data analysis. *npj Artificial Intelligence 2026 2:1* **2** (2026–01–16). https://doi.org/10.1038/s44387-025-00064-0
31 Rood, J. E., Hupalowska, A. & Regev, A. Toward a foundation model of causal cell and tissue biology with a Perturbation Cell and Tissue Atlas. *Cell* **187** (2024/08/22). https://doi.org/10.1016/j.cell.2024.07.035
32 Bunne, C. *et al.* How to build the virtual cell with artificial intelligence: Priorities and opportunities. *Cell* **187**, 7045–7063 (2024). https://doi.org/10.1016/j.cell.2024.11.015
33 Luo, E., Hao, M., Wei, L. & Zhang, X. scDiffusion: conditional generation of high-quality single-cell data using diffusion model. *Bioinformatics* **40** (2024/09/02). https://doi.org/10.1093/bioinformatics/btae518
34 Song, D. *et al.* scDesign3 generates realistic in silico data for multimodal single-cell and spatial omics. *Nature Biotechnology 2023 42:2* **42** (2023–05–11). https://doi.org/10.1038/s41587-023-01772-1
35 Levine, D. *et al.* Cell2Sentence: Teaching Large Language Models the Language of Biology. *bioRxiv* (2024 Oct 29). https://doi.org/10.1101/2023.09.11.557287

36 Kamimoto, K. *et al.* Dissecting cell identity via network inference and in silico gene perturbation. *Nature 2023 614:7949* **614** (2023–02–08). https://doi.org/10.1038/s41586-022-05688-9
37 Roohani, Y. *et al.* Predicting transcriptional outcomes of novel multigene perturbations with GEARS. *Nature Biotechnology 2023 42:6* **42** (2023–08–17). https://doi.org/10.1038/s41587-023-01905-6
38 Chuai, G. *et al.* Towards building a World Model to simulate perturbation-induced cellular dynamics by AlphaCell. *bioRxiv* (2026–03–05). https://doi.org/10.64898/2026.03.02.709176

## Figures

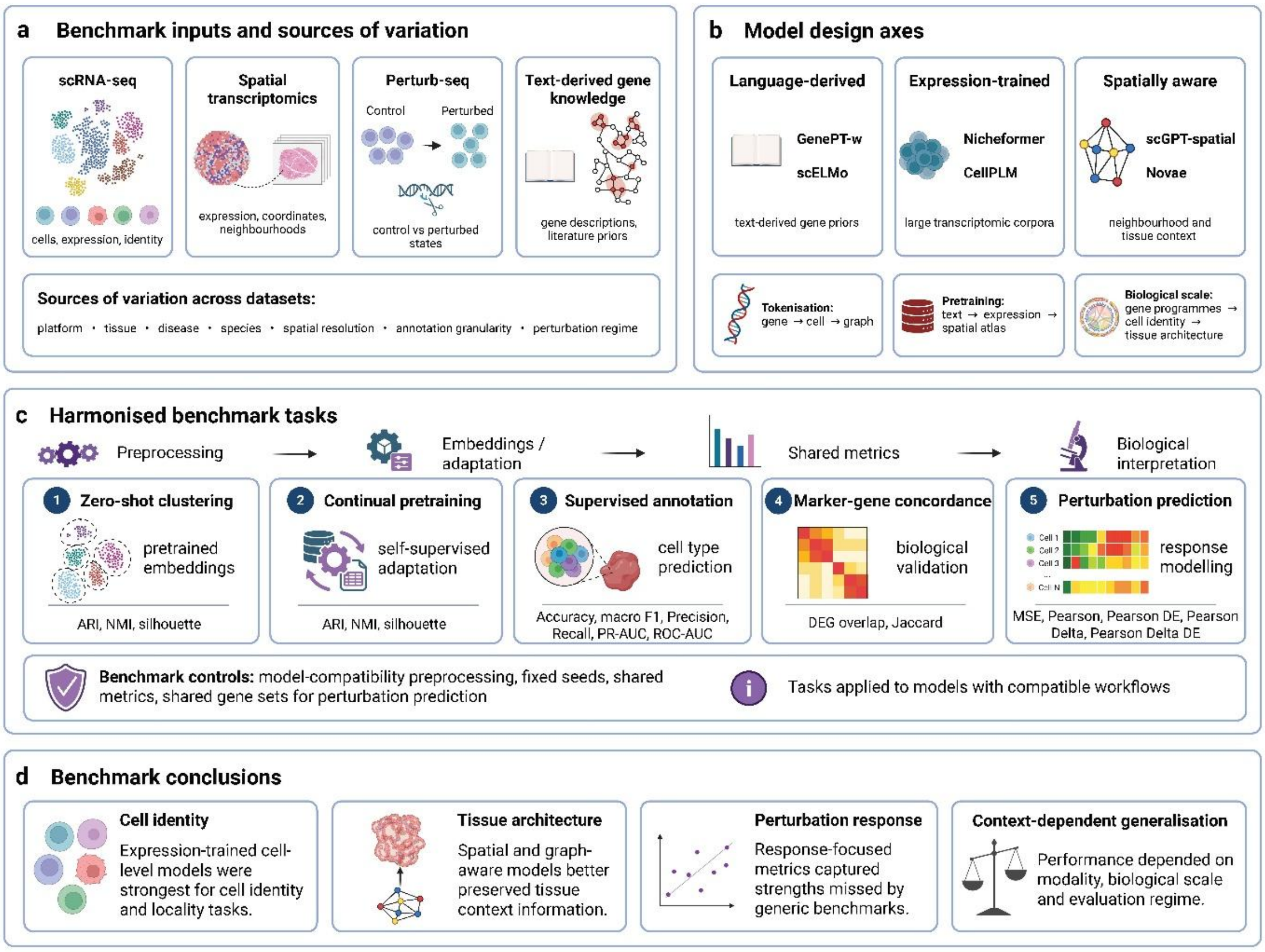

**Figure 1. Conceptual framework for benchmarking foundation-model generalisation across single-cell and spatial omics. a,** Overview of the biological inputs and sources of variation represented in the benchmark. The study evaluates models across scRNA-seq, spatial transcriptomics, Perturb-seq and text-derived gene knowledge. These inputs capture complementary biological information: dissociated cellular identity, tissue architecture and neighbourhood structure, perturbation-induced transcriptional responses, and literature-derived gene-function priors. The lower ribbon summarises biological and technical variation across datasets, including platform, tissue, disease, species, spatial resolution, annotation granularity and perturbation regime. **b,** Model design axes represented by the six evaluated foundation models. GenePT-w and scELMo represent gene- or language-derived priors; CellPLM and scGPT-spatial represent expression-trained cell-level models; and Nicheformer and Novae represent spatial or microenvironment-aware architectures. The lower boxes summarise how these models differ in tokenisation

unit, pretraining source and biological scale, from gene programmes to cell identity and tissue architecture. **c,** Harmonised benchmark workflow and downstream tasks. Models were evaluated using model-compatible preprocessing, embedding extraction or adaptation, shared metrics and biological interpretation. The benchmark spans zero-shot clustering, continual pretraining, supervised annotation, marker-gene concordance and perturbation prediction. Benchmark controls included fixed random seeds, shared metrics, shared test sets where applicable and shared gene sets for perturbation prediction; tasks were applied to models with compatible workflows. d, Summary of the main benchmark conclusions. Expression-trained cell-level models were strongest for many cell-identity tasks, spatial and graph-aware models better preserved tissue-scale organisation, and perturbation-response modelling required response-focused metrics beyond global expression reconstruction. Overall, model performance was context-dependent, with rankings shifting according to modality, biological scale, domain shift and evaluation metric.

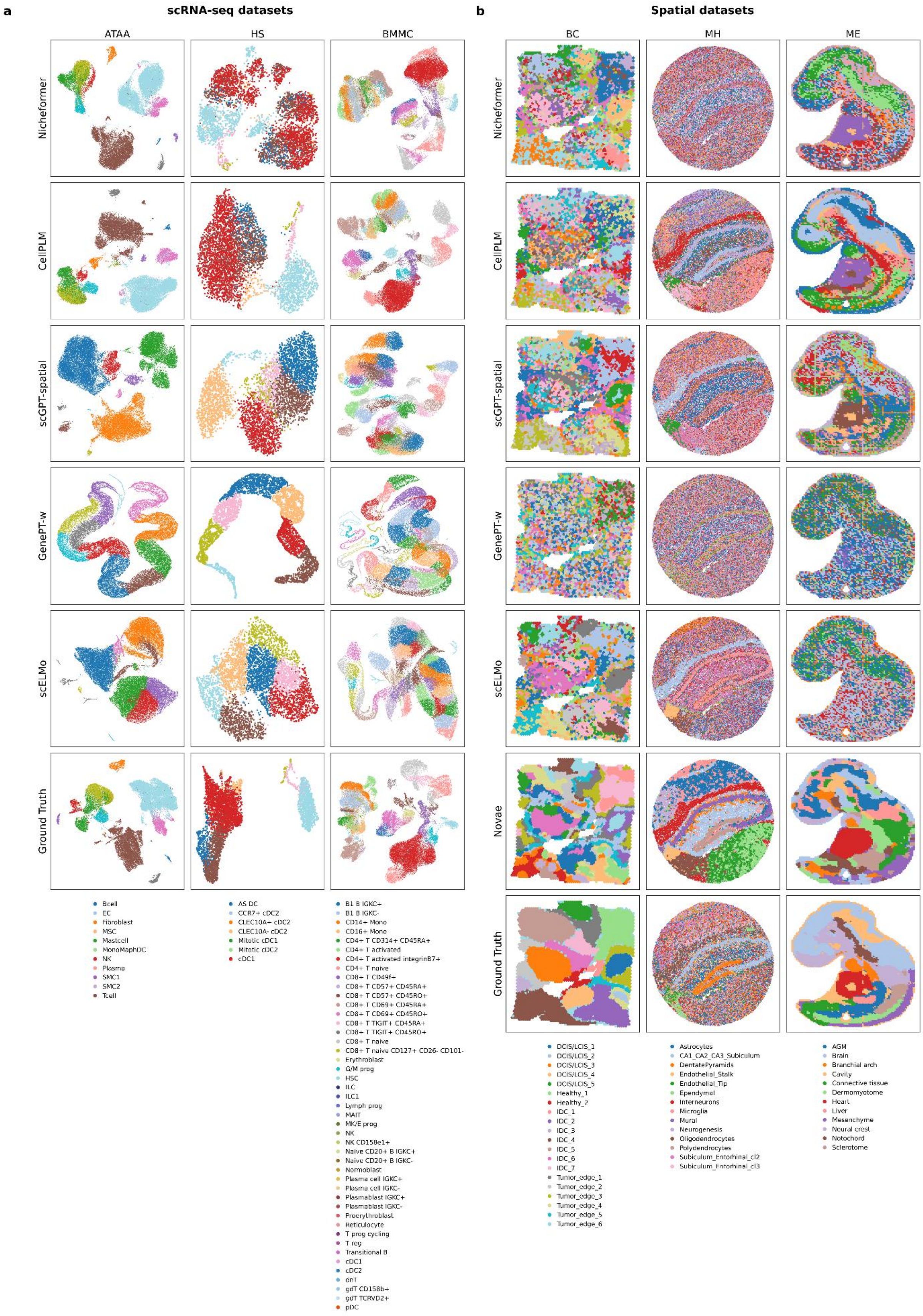

**Figure 2. Zero-shot foundation-model embeddings recover biological structure unevenly across single-cell and spatial modalities. a**, UMAP visualisations of zero-

shot embeddings from Nicheformer, CellPLM, scGPT-spatial, GenePT-w and scELMo across three scRNA-seq datasets: ascending thoracic aortic aneurysm (ATAA), human spleen dendritic cells (HS) and bone marrow mononuclear cells (BMMC). Points are coloured by ground-truth cell type labels; the bottom row shows the ground-truth reference visualisation for comparison. **b,** Spatial visualisations of zero-shot model-derived clusters or domains across three spatial transcriptomics datasets: human breast cancer (BC/HBCA1), mouse hippocampus (MH/MHPC) and mouse embryo (ME/MOSTA E9.5). The bottom row shows curated ground-truth spatial annotations. For spatial panels, colours represent model-derived domains or annotated domains within each plot and should be interpreted together with the quantitative ARI, NMI and silhouette scores in **Table 2**.

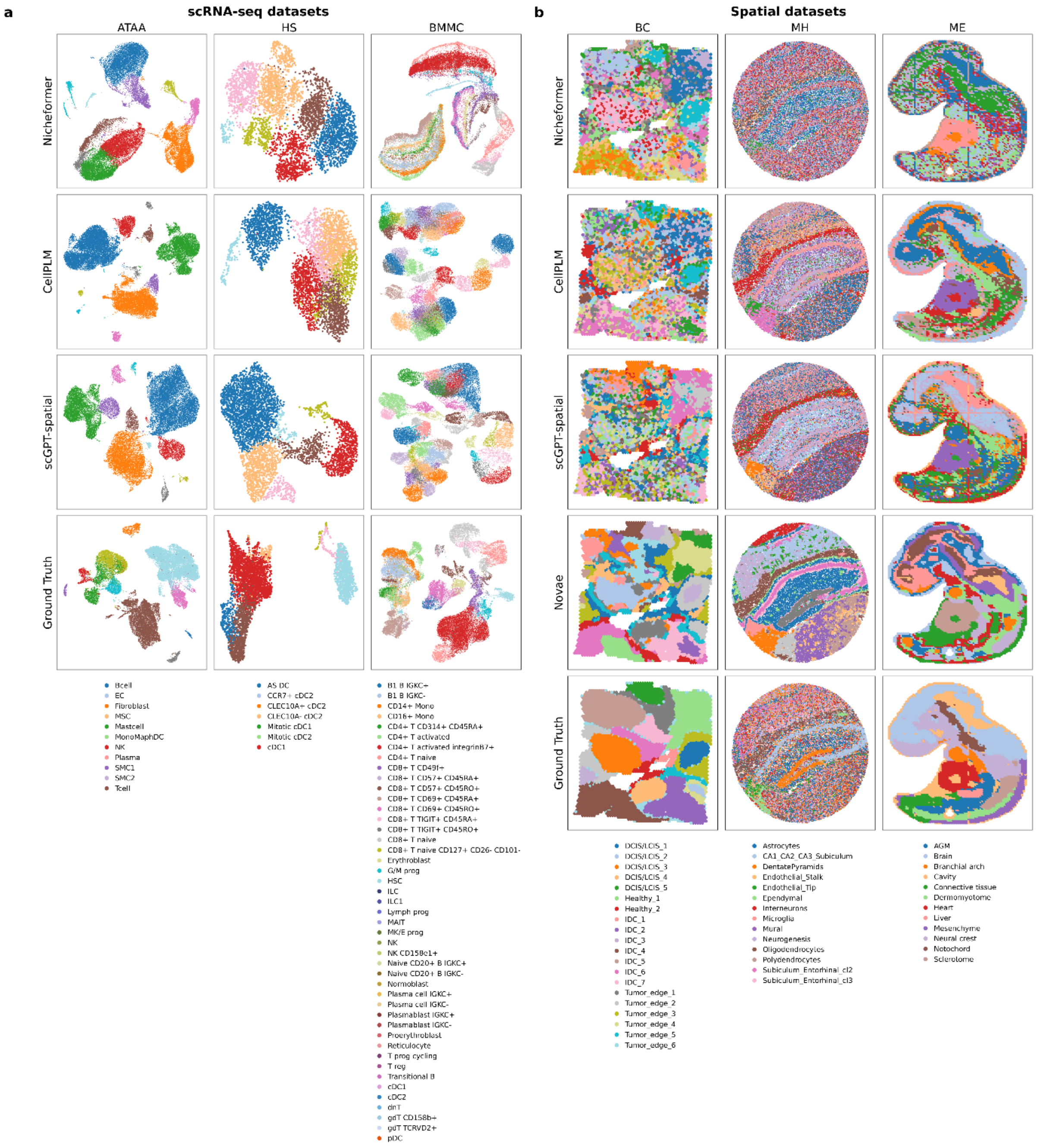

**Figure 3. Continual pretraining selectively refines cell-state and tissue-domain representations. a**, UMAP visualisations of embeddings from continually pretrained models for three scRNA-seq datasets: ascending thoracic aortic aneurysm (ATAA), human spleen dendritic cells (HS) and bone marrow mononuclear cells (BMMC). Points represent cells and are coloured by ground-truth cell type labels. The bottom row shows ground-truth visualisations for comparison. **b,** Spatial plots of continually pretrained model outputs for three spatial transcriptomics datasets: human breast cancer (BC/HBCA1), mouse hippocampus (MH/MHPC) and mouse embryo (ME/MOSTA E9.5). Model panels are coloured by predicted Leiden clusters or model-

derived spatial domains, whereas the bottom row shows curated ground-truth spatial annotations. Only models with supported or adapted continual-pretraining workflows are shown.

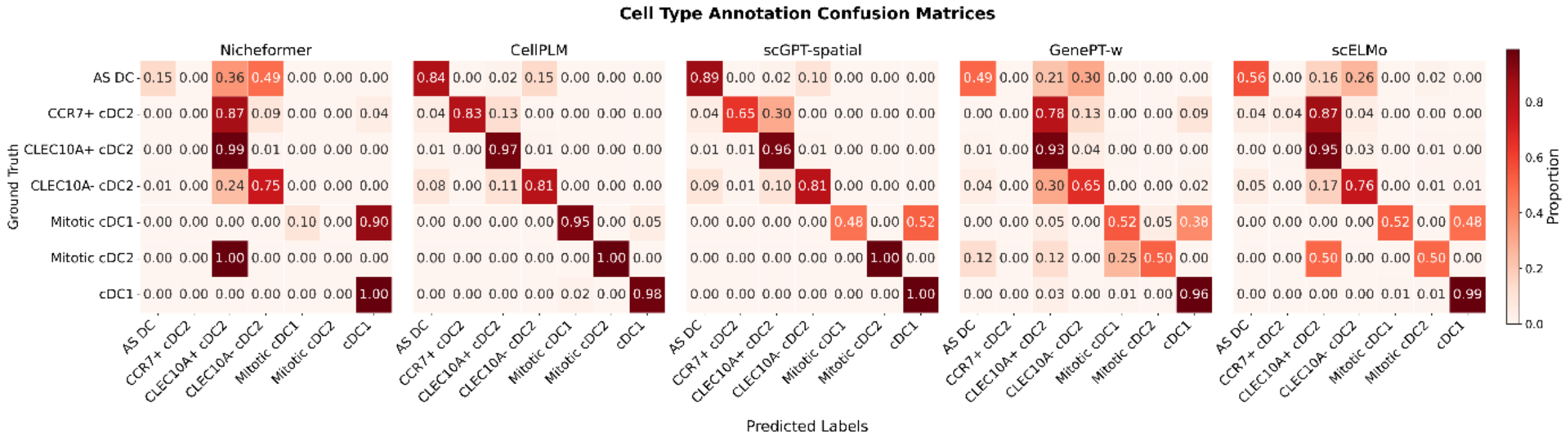

**Figure 4. Supervised annotation of closely related dendritic-cell states reveals class-specific model failures.** Row-normalised confusion matrices for supervised cell type annotation on the human spleen dendritic-cell dataset. Rows indicate ground-truth labels and columns indicate predicted labels for seven dendritic-cell states: AS DC, CCR7$^{+}$ cDC2, CLEC10A$^{+}$ cDC2, CLEC10A$^{-}$ cDC2, mitotic cDC1, mitotic cDC2 and cDC1. Each value represents the proportion of cells from a given ground-truth class assigned to each predicted class; darker red indicates a higher proportion. A strong diagonal therefore indicates high per-class annotation fidelity, whereas off-diagonal signal reveals systematic confusion between related dendritic-cell states. CellPLM shows the most consistent diagonal structure, including strong recovery of CCR7$^{+}$ cDC2, CLEC10A$^{+/-}$ cDC2, mitotic cDC1, mitotic cDC2 and cDC1. scGPT-spatial also performs strongly but shows residual confusion between CCR7$^{+}$ cDC2 and CLEC10A$^{+}$ cDC2, and between mitotic cDC1 and cDC1. Nicheformer, GenePT-w and scELMo recover dominant cDC1 and CLEC10A$^{+}$ cDC2 populations but show broader off-diagonal errors, particularly for AS DC, CCR7$^{+}$ cDC2 and mitotic DC states.

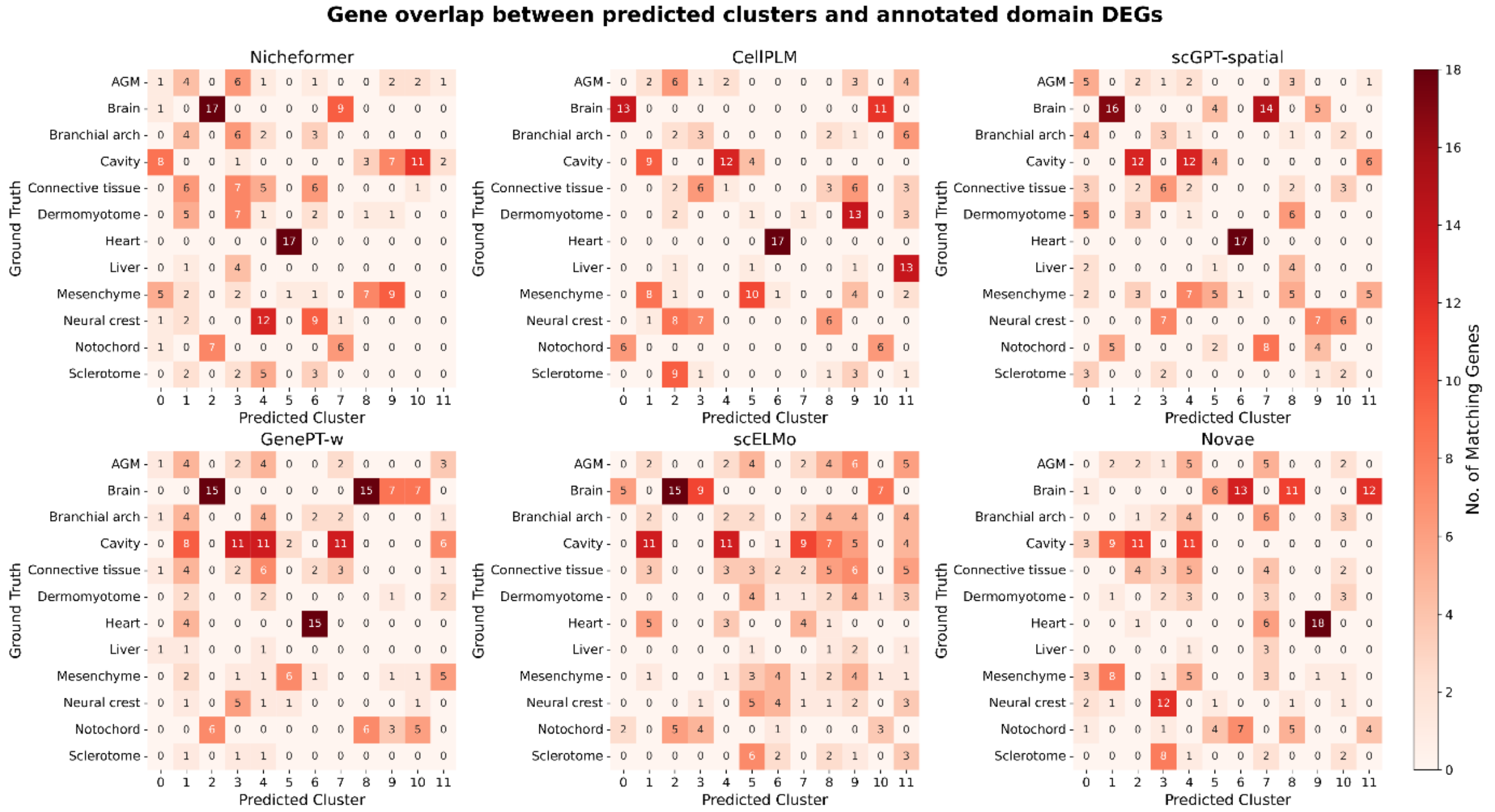

**Figure 5. Marker-gene concordance between model-derived spatial clusters and annotated developmental domains in the MOSTA E9.5 mouse embryo.** Heatmaps show raw overlap counts between the top 20 differentially expressed genes identified for each annotated anatomical domain and the top 20 differentially expressed genes identified for each model-derived cluster. Rows correspond to ground-truth embryo domains, including aorta–gonad–mesonephros (AGM), brain, branchial arch, cavity, connective tissue, dermomyotome, heart, liver, mesenchyme, neural crest, notochord and sclerotome. Columns correspond to predicted clusters, with cluster IDs assigned independently for each model and therefore not directly comparable across models. Colour intensity and numeric values indicate the number of shared marker genes, with darker red denoting greater overlap. Strong row-specific peaks indicate that a predicted cluster recovers a biologically interpretable developmental programme, whereas distributed signal across multiple clusters suggests shared lineage programmes, spatial gradients or incomplete separation of neighbouring embryonic domains. Corresponding Jaccard-index heatmaps showing normalised marker-gene overlap are provided in **Supplementary Figure S2**.

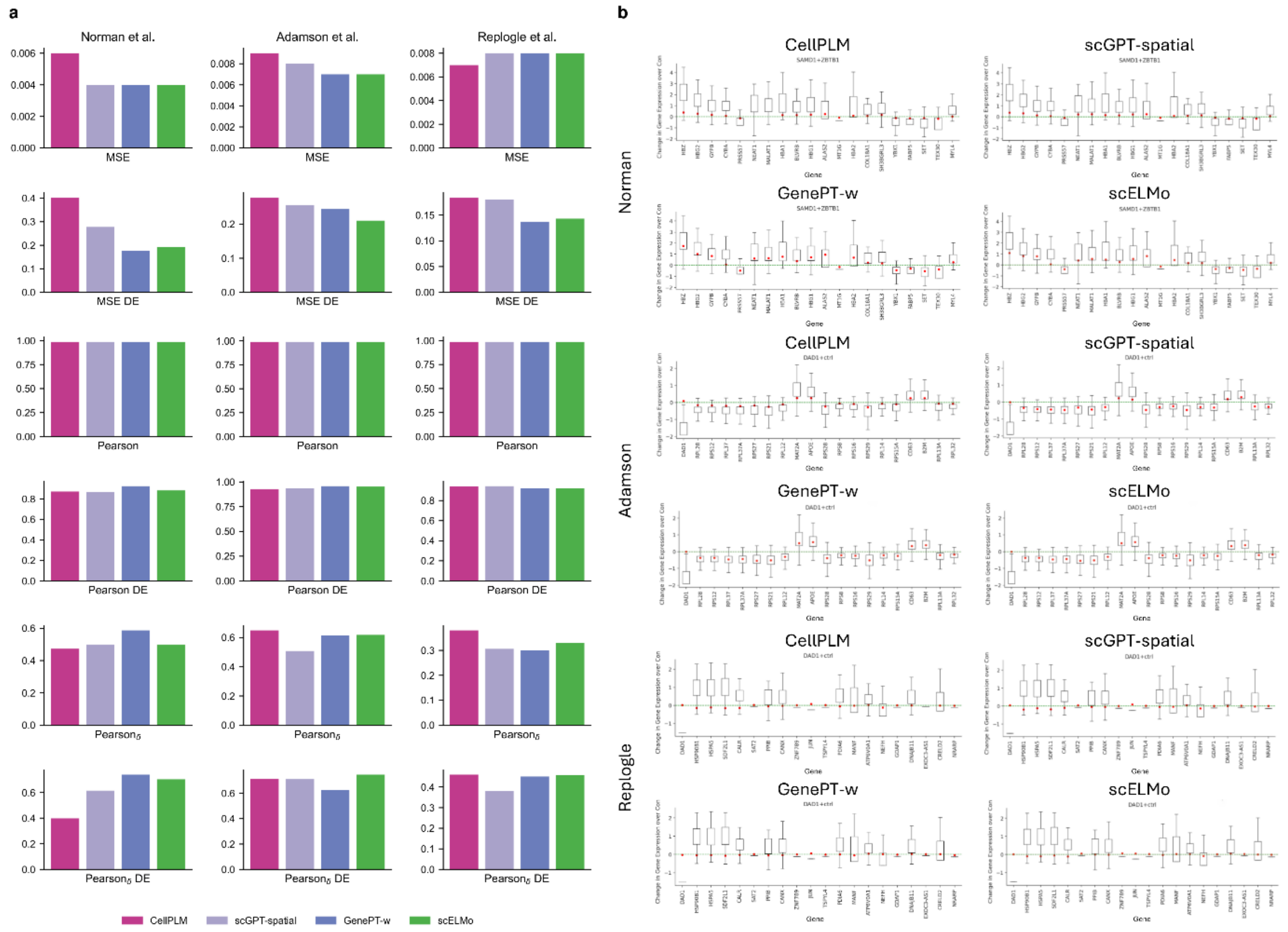

**Figure 6. Perturbation prediction separates global expression reconstruction from perturbation-response modelling. a**, Bar plots showing perturbation-prediction performance for CellPLM, scGPT-spatial, GenePT-w and scELMo across three Perturb-seq datasets: Adamson, Norman and Replogle. Performance is reported using MSE and Pearson correlation across all evaluated genes, the same metrics restricted to top differentially expressed genes, and delta-based Pearson metrics measuring perturbation-induced expression change relative to control. Lower values indicate better performance for MSE and MSE DE, whereas higher values indicate better performance for Pearson, Pearson DE, Pearson δ and Pearson δ DE. All models were evaluated on the same gene set within each dataset. **b**, Representative held-out perturbation examples showing observed and predicted gene-expression changes relative to control for selected response genes. Boxplots show the observed distribution of gene-expression change across cells, red points indicate model-predicted changes, and the green dashed line marks no change relative to control.

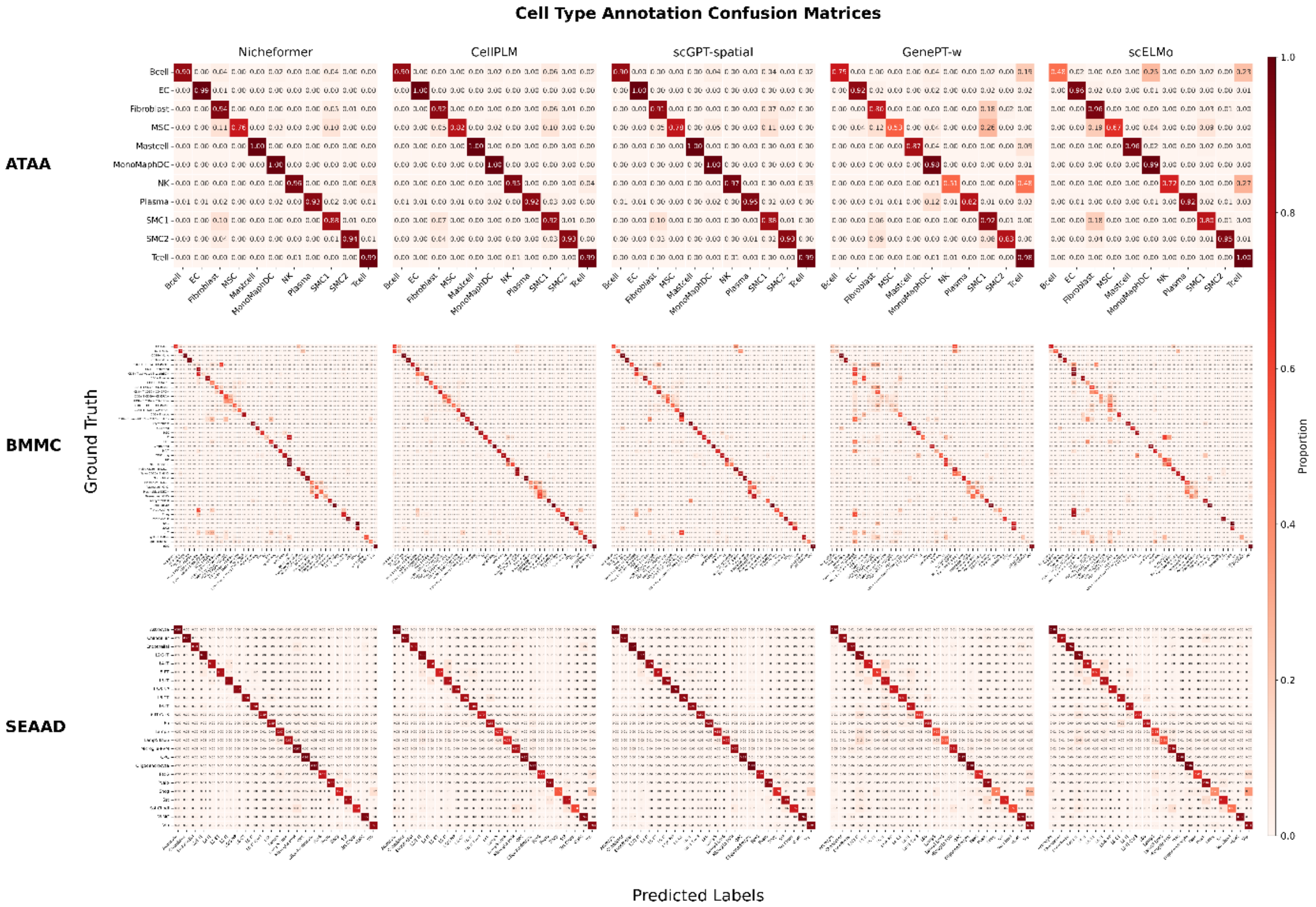

**Supplementary Figure S1.** Confusion heatmaps showing similarity between test set ground truth and predicted cell type labels for cell type annotation task (ATAA, BMMC and SEAAD datasets).

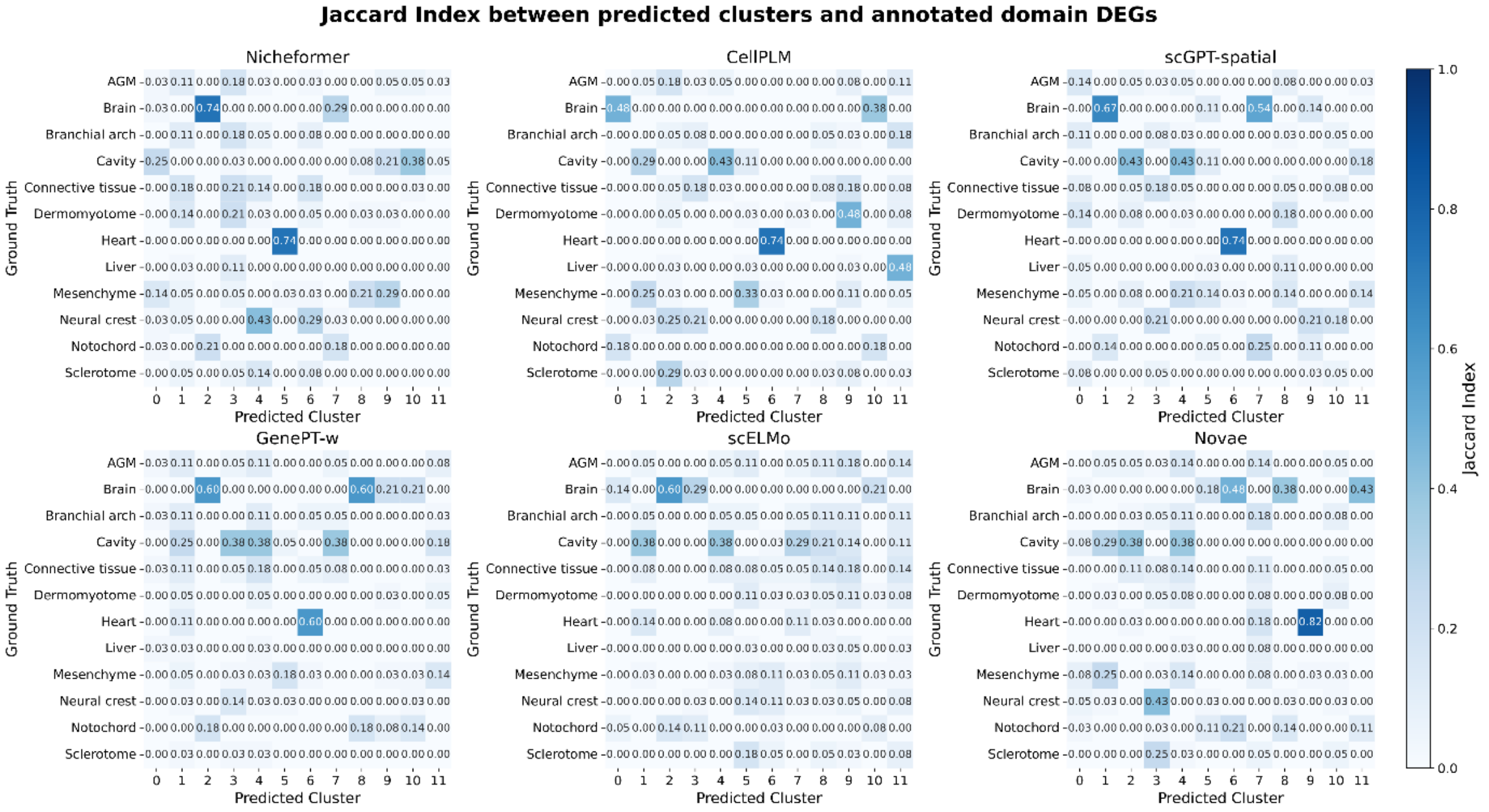

**Supplementary Figure S2.** Jaccard-index heatmaps showing normalised overlap between top-20 DEGs from annotated MOSTA E9.5 domains and model-derived clusters.

## Tables

**Table 1. Overview of foundation models included in the benchmark.** Models are compared by tokenisation strategy, pretraining data, modality, architecture, spatial awareness, reported tasks, pretraining objective and publication.

| Model | Tokenisation Strategy | Pretraining Data | Modality | Architecture | Spatial Awareness | Reported Tasks | Pretraining Objective | Publication |
|---|---|---|---|---|---|---|---|---|
| Nicheformer | Gene rank-based tokens | 110 million cells (57M scRNA-seq + 53M ST), human and mouse | Single-cell + spatial transcriptomics | Transformer encoder | Implicit (pretraining on spatial data) | Spatial label/niche/region prediction, neighbourhood composition prediction, neighbourhood density prediction | Masked gene expression prediction | Nature Methods (2025) |
| CellPLM | Cell-level tokens | 9M scRNA-seq + 2M ST cells, human | Single-cell + spatial transcriptomics | Transformer-based encoder + Gaussian Mixture VAE + Decoder | No | Zero-shot clustering, scRNA-seq denoising, spatial transcriptomic imputation, cell type annotation, perturbation prediction | Masked gene expression prediction | ICLR 2024 |
| scGPT-spatial | Gene-level tokens | 30M ST cells/spots, human | Single-cell + spatial transcriptomics | scGPT backbone + MoE decoder | Implicit (spatially-aware sampling) | Spatial gene imputation, cell-type deconvolution, spatial domain clustering, data integration | Masked gene expression prediction | bioRxiv |
| GenePT | Text-derived gene embeddings (expression-weighted aggregation) | No pretraining | Single-cell + text | Transformer decoder (GPT) | No | Gene functionality prediction, gene property prediction, GGI prediction, PPI prediction, gene program derivation, cell type annotation, disease phenotype prediction, batch integration | N/A | Nature Biomedical Engineering (2025) |
| scELMo | Text-derived gene embeddings (expression-weighted aggregation) | No pretraining | Single-cell + text | Transformer decoder (GPT) | No | Zero-shot clustering, cell type annotation, perturbation analysis, in-silico treatment analysis, batch effect correction | N/A | Patterns (2026) |
| Novae | Cells as nodes and gene expression vectors as features | 29 million ST cells | Spatial transcriptomics | Graph attention encoder | Explicit (graph-based architecture) | Zero-shot spatial domain inference, spatial domain trajectory analysis, spatially variable gene analysis, spatial pathway analysis, batch effect correction | SwAV (Contrastive learning + clustering) | Nature Methods (2025) |

**Table 2.** Zero-shot cell type clustering results across three scRNA-seq and three spatial transcriptomics datasets. Performance was assessed using ARI, NMI and Silhouette scores.

| Model | Dataset | ARI | NMI | Silhouette |
|---|---|---|---|---|
| | | **scRNA-seq** | | |
| Nicheformer | ATAA | 0.62 | 0.78 | 0.21 |
| | HS | 0.21 | 0.34 | 0.06 |
| | BMMC | 0.25 | 0.64 | 0.07 |
| CellPLM | ATAA | 0.82 | 0.80 | 0.26 |
| | HS | 0.41 | 0.52 | 0.14 |
| | BMMC | 0.31 | 0.69 | 0.13 |
| scGPT-spatial | ATAA | 0.87 | 0.85 | 0.30 |
| | HS | 0.47 | 0.56 | 0.15 |
| | BMMC | 0.27 | 0.63 | 0.14 |
| GenePT-w | ATAA | 0.50 | 0.65 | 0.10 |
| | HS | 0.21 | 0.34 | 0.14 |
| | BMMC | 0.12 | 0.37 | -0.26 |
| scELMo | ATAA | 0.72 | 0.71 | 0.18 |
| | HS | 0.33 | 0.41 | 0.27 |
| | BMMC | 0.50 | 0.57 | 0.19 |
| | | **Spatial Transcriptomics** | | |
| Nicheformer | BC | 0.27 | 0.42 | 0.05 |
| | MH | 0.02 | 0.05 | 0.06 |
| | ME | 0.20 | 0.33 | 0.05 |
| CellPLM | BC | 0.17 | 0.32 | 0.09 |
| | MH | 0.13 | 0.20 | 0.06 |
| | ME | 0.26 | 0.39 | 0.10 |
| scGPT-spatial | BC | 0.28 | 0.45 | 0.09 |
| | MH | 0.12 | 0.19 | 0.07 |
| | ME | 0.19 | 0.30 | 0.08 |
| GenePT-w | BC | 0.08 | 0.24 | -0.35 |
| | MH | 0.01 | 0.03 | -0.26 |
| | ME | 0.03 | 0.10 | -0.29 |
| scELMo | BC | 0.35 | 0.44 | 0.18 |
| | MH | 0.10 | 0.15 | 0.30 |
| | ME | 0.08 | 0.14 | 0.31 |
| Novae | BC | 0.25 | 0.46 | 0.19 |
| | MH | 0.07 | 0.13 | 0.15 |
| | ME | 0.22 | 0.39 | 0.13 |

**Table 3.** Continually pretrained cell type clustering results across three scRNA-seq and three spatial transcriptomics datasets. Performance was assessed using ARI, NMI and Silhouette scores.

| Model | Dataset | ARI | NMI | Silhouette |
|---|---|---|---|---|
| | | **scRNA-seq** | | |
| Nicheformer | ATAA | 0.57 | 0.74 | 0.26 |
| | HS | 0.14 | 0.23 | 0.04 |
| | BMMC | 0.23 | 0.59 | 0.08 |
| CellPLM | ATAA | 0.84 | 0.82 | 0.26 |
| | HS | 0.37 | 0.52 | 0.14 |
| | BMMC | 0.28 | 0.66 | 0.12 |
| scGPT-spatial | ATAA | 0.87 | 0.85 | 0.18 |
| | HS | 0.68 | 0.62 | 0.34 |
| | BMMC | 0.28 | 0.66 | 0.19 |
| | | **Spatial Transcriptomics** | | |
| Nicheformer | BC | 0.26 | 0.41 | 0.04 |
| | MH | 0.03 | 0.10 | 0.04 |
| | ME | 0.19 | 0.32 | 0.04 |
| CellPLM | BC | 0.16 | 0.31 | 0.10 |
| | MH | 0.15 | 0.21 | 0.06 |
| | ME | 0.27 | 0.40 | 0.09 |
| scGPT-spatial | BC | 0.17 | 0.33 | 0.10 |
| | MH | 0.11 | 0.21 | 0.15 |
| | ME | 0.21 | 0.36 | 0.17 |
| Novae | BC | 0.31 | 0.51 | 0.17 |
| | MH | 0.07 | 0.14 | 0.32 |
| | ME | 0.18 | 0.35 | 0.20 |

**Table 4**. Cell type annotation performance across the test sets of four datasets, evaluated using accuracy, precision, recall, macro-F1, PR-AUC and ROC-AUC metrics.

| Model | Dataset | Accuracy | Precision | Recall | Macro-F1 | PR-AUC | ROC-AUC |
|---|---|---|---|---|---|---|---|
| Nicheformer | ATAA | 0.97 | 0.97 | 0.97 | 0.94 | 0.98 | 0.99 |
| | HS | 0.83 | 0.83 | 0.83 | 0.43 | 0.81 | 0.97 |
| | BMMC | 0.84 | 0.84 | 0.84 | 0.61 | 0.72 | 0.99 |
| | SEAAD | 0.94 | 0.94 | 0.94 | 0.88 | 0.95 | 0.99 |
| CellPLM | ATAA | 0.97 | 0.97 | 0.97 | 0.95 | 0.98 | 0.99 |
| | HS | 0.93 | 0.93 | 0.93 | 0.88 | 0.96 | 0.99 |
| | BMMC | 0.86 | 0.86 | 0.86 | 0.73 | 0.80 | 0.99 |
| | SEAAD | 0.91 | 0.91 | 0.91 | 0.84 | 0.91 | 0.99 |
| scGPT-spatial | ATAA | 0.97 | 0.95 | 0.94 | 0.94 | 0.98 | 0.99 |
| | HS | 0.91 | 0.89 | 0.83 | 0.84 | 0.96 | 0.99 |
| | BMMC | 0.84 | 0.67 | 0.66 | 0.65 | 0.91 | 0.99 |
| | SEAAD | 0.94 | 0.89 | 0.89 | 0.89 | 0.97 | 0.99 |
| GenePT-w | ATAA | 0.92 | 0.92 | 0.81 | 0.85 | 0.87 | 0.96 |
| | HS | 0.81 | 0.67 | 0.58 | 0.61 | 0.66 | 0.89 |
| | BMMC | 0.71 | 0.48 | 0.44 | 0.45 | 0.43 | 0.85 |
| | SEAAD | 0.90 | 0.85 | 0.77 | 0.79 | 0.86 | 0.96 |
| scELMo | ATAA | 0.95 | 0.96 | 0.86 | 0.89 | 0.95 | 0.99 |
| | HS | 0.86 | 0.71 | 0.62 | 0.64 | 0.71 | 0.99 |
| | BMMC | 0.80 | 0.58 | 0.51 | 0.51 | 0.61 | 0.86 |
| | SEAAD | 0.88 | 0.82 | 0.77 | 0.79 | 0.78 | 0.93 |
| Novae | SEAAD | 0.30 | 0.09 | 0.05 | 0.03 | 0.04 | 0.50 |

**Table 5.** Perturbation prediction benchmarking results across three Perturb-seq datasets. Models were evaluated using MSE, MSE DE, Pearson, Pearson DE, Pearson Delta and Pearson Delta DE metrics.

| Model | Dataset | MSE | MSE DE | Pearson | Pearson DE | Pearson Delta | Pearson Delta DE |
|---|---|---|---|---|---|---|---|
| CellPLM | Norman | 0.0064 | 0.4034 | 0.9826 | 0.8724 | 0.4727 | 0.3982 |
| | Adamson | 0.0091 | 0.2778 | 0.9876 | 0.9236 | 0.6487 | 0.7112 |
| | Replogle | 0.0068 | 0.1844 | 0.9868 | 0.9406 | 0.3801 | 0.4604 |
| scGPT-spatial | Norman | 0.0041 | 0.2784 | 0.9875 | 0.8648 | 0.5002 | 0.6146 |
| | Adamson | 0.0080 | 0.2552 | 0.9872 | 0.9369 | 0.5080 | 0.7097 |
| | Replogle | 0.0076 | 0.1804 | 0.9863 | 0.9437 | 0.3057 | 0.3801 |
| GenePT-w | Norman | 0.0043 | 0.1757 | 0.9879 | 0.9252 | 0.5871 | 0.7389 |
| | Adamson | 0.0068 | 0.2441 | 0.9896 | 0.9576 | 0.6144 | 0.6238 |
| | Replogle | 0.0076 | 0.1365 | 0.9868 | 0.9248 | 0.3001 | 0.4505 |
| scELMo | Norman | 0.0043 | 0.1913 | 0.9880 | 0.8818 | 0.5004 | 0.7032 |
| | Adamson | 0.0073 | 0.2104 | 0.9883 | 0.9524 | 0.6178 | 0.7450 |
| | Replogle | 0.0077 | 0.1426 | 0.9866 | 0.9214 | 0.3287 | 0.4579 |

**Supplementary Table S1.** Models considered but not included in the primary benchmark.

| Model/tool | Primary purpose | Reason not included in primary benchmark |
|---|---|---|
| scDiffusion[33] | Conditional generation of synthetic scRNA-seq data | Generative simulator rather than cross-task embedding foundation model |
| scDesign3[34] | Statistical simulation of single-cell/spatial multi-omics | Data simulator, not pretrained representation model |
| Cell2Sentence[35] | Language-style single-cell representation | Relevant but overlaps with language/text axis already represented by GenePT-w and scELMo; not included in current cross-modality panel |
| CellOracle[36] | GRN inference and in silico perturbation | Mechanistic regulatory-network tool rather than general foundation representation model |
| GEARS[37] | Perturbation prediction | Used as perturbation framework/decoder; should be perturbation baseline, not primary FM |
| AlphaCell[38] | Ambiguous / model not verified at benchmark freeze | Not included without citable, reproducible, comparable implementation |

**Supplementary Table S2.** Zero-shot clustering metrics (k-means).

| Model | Modality | Dataset | ARI | NMI | Silhouette |
|---|---|---|---|---|---|
| GenePT-w | scRNA-seq | ATAA | 0.16 | 0.22 | -0.13 |
| | | HS | 0.08 | 0.15 | 0.13 |
| | | BMMC | 0.07 | 0.19 | -0.14 |
| | Spatial Transcriptomics | BC | 0.14 | 0.24 | -0.36 |
| | | MH | 0.01 | 0.03 | -0.50 |
| | | ME | 0.06 | 0.13 | -0.26 |